\documentclass[12pt,letterpaper]{article}
\pdfoutput=1

\usepackage{graphicx,array}
\usepackage{url}
\usepackage{color}
\usepackage{latexsym}
\usepackage{amsthm}
\usepackage{amsmath}
\usepackage{amssymb}
\usepackage{amsfonts}
\usepackage[numbers,sort&compress]{natbib}
\usepackage{bm}
\usepackage{slashed}
\usepackage{mathrsfs}
\usepackage{enumerate}
\usepackage{tikz}
\usepackage{siunitx}
\usepackage{mdframed}
\usepackage{setspace}

\usepackage[utf8x]{inputenc}%
\usepackage{tcolorbox}%

\usepackage{hyperref} 
\hypersetup{
    colorlinks=true,       
    linkcolor=red,          
    citecolor=blue,        
    filecolor=magenta,      
    urlcolor=blue           
}
\usepackage[all]{hypcap} 

\usepackage{natbib}
\usepackage[normalem]{ulem} 

\newcommand{\nc}{\newcommand}

\nc{\beq}{\begin{equation}}  
\nc{\eeq}{\end{equation}}  
\nc{\beqa}{\begin{eqnarray}}  
\nc{\eeqa}{\end{eqnarray}}  
\nc{\bit}{\begin{itemize}}  
\nc{\eit}{\end{itemize}}  

\def\GeV{\mathrm{GeV}}     

\newcommand{\eg}{{\it e.g.}}
\newcommand{\ie}{{\it i.e.}}
\newcommand{\Mpl}{M_{\rm pl}}

\newcommand{\rb}{\bar{r}}
\newcommand{\mwd}{m_{W'}}
\newcommand{\gstarc}{g_*^c}

\newcommand{\cI}{\mathcal{I}}

\newcommand{\longeq}{\scalebox{3}[1]{=}} 

\usepackage{floatrow}
\newfloatcommand{capbtabbox}{table}[][\FBwidth]

\usepackage{blindtext}

\DeclareRobustCommand\encircle[1]{%
  \tikz[baseline=(X.base)] 
    \node (X) [draw, shape=circle, inner sep=-0.7] {\strut \raisebox{-0.5pt}[0pt]{#1}};}
\newcommand{\Mcirc}{\encircle{\textbf{M}}}
\newcommand{\Mmono}{M_{\tiny \Mcirc}}
\newcommand{\Rmono}{R_{\tiny \Mcirc}}

\title{ \bf
Electroweak-Symmetric Dark Monopoles \\
 from Preheating
\author{\large Yang Bai, Mrunal Korwar, and Nicholas Orlofsky}
\date{\small \it 
Department of Physics, University of Wisconsin-Madison, Madison, WI 53706, USA
}
}

\begin{document}

\maketitle

\setlength{\parskip}{0.2ex}

\begin{abstract}	
If the dark sector contains 't Hooft-Polyakov monopoles and a small enough dark gauge coupling, dark monopoles could be a macroscopic dark matter candidate. Its Higgs-portal coupling to the Standard Model can modify the electroweak vacuum in the monopole interior. In the most striking cases, dark monopoles could even contain electroweak-symmetric cores and generate multi-hit signals at large-volume detectors. If they are produced via parametric resonance in the early Universe, monopoles with radii up to one micron and masses up to ten kilotonnes could account for all of dark matter.
\end{abstract}

\thispagestyle{empty}  
\newpage  
  
\setcounter{page}{1}  

\begingroup
\hypersetup{linkcolor=black,linktocpage}
\tableofcontents
\endgroup

\newpage

\section{Introduction}\label{sec:Introduction}

The observational evidence for dark matter (DM) indicates the high likelihood of a new dark particle or a dark sector of particles.  This dark sector may have a zoo of particles and gauge symmetries akin to the rich phenomenology of the Standard Model (SM).  If the dark sector's gauge symmetry breaking contains non-trivial homotopies, then stable topological defects charged under the dark gauge groups may form.  In some circumstances, these dark topological defects could even provide the dominant contribution to the DM energy density.  This paper will focus on the case where these dark topological defects are 't Hooft-Polyakov monopoles~\cite{tHooft:1974kcl,Polyakov:1974ek}.
Dark monopoles as dark matter have previously been studied in Refs.~\cite{Murayama:2009nj,Evslin:2012fe,Baek:2013dwa,Khoze:2014woa,Terning:2019bhg}.

The monopole structure gives it distinct properties compared to particle DM.  There is significant freedom for the dark monopole properties depending on the dark gauge symmetry breaking scale $f$ and dark gauge coupling $g$.  In particular, its mass and radius are $\Mmono \simeq 4 \pi f / g$ and $\Rmono \simeq 1/(g\,f)$.  If $g$ is very small, the mass and radius can both be macroscopically large, making monopoles a macroscopic DM candidate \cite{Jacobs:2014yca}. This is in contrast to the nontopological soliton models~\cite{Friedberg:1976me,Coleman:1985ki}, where an accumulation of a large amount of global charges is needed to increase the radius.

At the renormalizable level, the only interactions between the dark and visible sectors come from the quartic coupling  between two Higgs and two dark Higgs fields. Tests for the possibility of dark monopoles as DM candidates rely significantly on this coupling.  In this paper, we point out that if the dark sector couples to the SM via a Higgs portal, then the properties of the electroweak (EW) vacuum may be modified in the interior of dark topological defects, which can lead to non-trivial interactions with SM particles.  In the most striking cases with a large dark monopole radius, the EW symmetry may be fully restored on length scales significantly longer than the EW scale.  Such large-radius EW-symmetric dark monopoles will have similar phenomenological signals to EW-symmetric dark matter balls \cite{Ponton:2019hux}, which can form bound states with nuclei and lead to multi-hit signals in direct detection experiments \cite{Bai:2019ogh}.  Because we are interested in exploring effects on the EW vacuum structure, we will give particular emphasis to EW-symmetric monopoles with large radius compared to the EW length scale. For the cases without EW restoration, the interaction of (even point-like) monopoles with the Higgs boson is enhanced, simply because the monopole should be treated as a coherent composite state of the fundamental scalar coupling to the Higgs. So, traditional DM direct detection experiments have the potential to discover dark monopoles with a weak Higgs-portal coupling.

The formation of monopoles during thermal cosmological phase transitions has been detailed by Kibble~\cite{Kibble:1976sj} and Zurek~\cite{Zurek:1985qw}, and their relic abundance following plasma-assisted annihilations was estimated by Preskill~\cite{Preskill:1979zi}.  As we will see, these well-known mechanisms do allow for the production of EW-symmetric dark monopoles.  However, if the dark monopoles are to make up the majority of DM, the plasma-assisted annihilations effectively fix the dark monopole radius to a particular value (regardless of monopole mass, dark symmetry-breaking scale, or dark gauge coupling).  Thus, dark monopoles must be point like compared to the EW scale to account for all of DM.  
Dark monopoles with radii larger than the EW scale can still be produced, but after they have annihilated they will only account for a small fraction of DM.  Even so, we will show that such a subdominant population of large-radius EW-symmetric dark monopoles may still be phenomenologically interesting.  

Another possibility that we will explore in this paper is dark monopole production during preheating~\cite{Dolgov:1989us,Traschen:1990sw,Kofman:1994rk,Kofman:1997yn}.  Unlike the case of monopole production during a cosmological phase transition, in this case the gauge-symmetry breaking occurs before the end of inflation.  Then, the inflaton oscillations coupled to the dark sector induce a parametric resonance effect that enhances field fluctuations in the gauge-symmetry-breaking field, leading to monopole formation (this has been explored for other topological defects in Refs.~\cite{Kasuya:1997ha,Khlebnikov:1998sz,Kasuya:1998td,Tkachev:1998dc,Kasuya:1999hy,Rajantie:2000fd,Kawasaki:2013iha}). Different from the thermal phase transition case, the dark charged states could be non-relativistic and have negligible abundance to induce monopole-antimonopole annihilations. We will show that this expands the possible radii for dark monopoles to comprise all of DM.

In the next section, we will consider the simplest realization of a monopole coupled to the SM Higgs.  We will show how the EW vacuum expectation value (VEV) is modified in the interior of monopoles, including the possibility that the EW symmetry is restored.  Then, in Sec.~\ref{sec:kibble-zurek}, we will review the cosmological phase transition production and plasma-assisted annihilations of monopoles.  Next, we will discuss monopole production during preheating and show the allowed dark monopole masses and radii in Sec.~\ref{sec:preheating}.  In Sec.~\ref{sec:pheno}, we will discuss the phenomenological implications, including direct detection of dark monopoles.  Finally, we conclude in Sec.~\ref{sec:conclusion}.

\section{Dark monopole model}\label{sec:simple}

\subsection{The 't Hooft-Polyakov monopole}
If a non-trivial second homotopy is present in the theory vacuum manifold of the dark sector, it provides for the existence of 't Hooft-Polyakov monopoles~\cite{tHooft:1974kcl,Polyakov:1974ek}. A topologically non-trivial field configuration with a conserved magnetic charge under an unbroken $U(1)$ can be realized by a spontaneously broken gauge theory with coset space $G/U(1)$ satisfying $\pi_2[G/U(1)] = \pi_1[U(1)] = \mathbb{Z}$. In this section, we review the simple $SU(2)/U(1)$ model and develop some notation used throughout the paper.  While this simple model is used for analysis throughout this paper, we expect the arguments to generalize to other coset spaces that admit monopoles.

To realize a dark monopole based on $\pi_2[SU(2)/U(1)]  = \mathbb{Z}$, we introduce a non-Abelian $SU(2)$ gauge theory in the dark sector. We adopt the simple Higgs mechanism by introducing a $SU(2)$ triplet scalar $\Phi^a$ with a potential that spontaneously breaks the symmetry at the scale $f$. 
The Lagrangian in the dark sector is 
\beq
\label{eq:basic-dark-lag}
\mathcal{L}_{\rm dark}= \frac{1}{2}\,(D_{\mu}\Phi)^{2} - \frac{1}{4}\mbox{Tr}(F_{\mu\nu}F^{\mu\nu}) - \frac{\lambda}{4} \left(|\Phi|^2 - f^2\right)^{2} \; ,
\eeq
where $D_{\mu}\Phi^{a} = \partial_{\mu}\Phi^{a} + g\, \epsilon^{abc}A^{b}_{\mu}\Phi^{c}$, $F_{\mu\nu}^{a}=\partial_{\mu}A_{\nu}^{a}-\partial_{\nu}A_{\mu}^{a} + g\, \epsilon^{abc}\,A^{b}_{\mu}A^{c}_{\nu}$, $|\Phi|^2 \equiv \sum_a (\Phi^a)^2$, $A_\mu^a$ is the gauge field, $a=1,2,3$ is the gauge index and $g$ is the gauge coupling. After symmetry breaking, two gauge bosons, $W'^\pm$, become massive with $\mwd = g\,f$. The remaining gauge boson stays massless and plays the role of dark photon. The dark Higgs boson in this model has mass $m_{h'} = \sqrt{2 \lambda}\,f$.  

The monopole is a time independent, finite energy, spherical solitonic solution for this system.  We use the ansatz corresponding to the so-called ``hedgehog gauge,'' with $A_0^a = 0$ and
\beqa
\label{eq:fieldRedef}
\Phi^{a} = \hat{r}^{a}\,f\,\phi(r) \,, \qquad  A^{a}_{i} =\frac{1}{g}\, \epsilon^{a i j}\, \hat{r}^{j}\,\Big{(} \dfrac{1 - u(r)}{r}\Big{)} \; .
\eeqa
Here, $\hat{r} \equiv \vec{r}/r$, and $\phi(r), u(r)$ are dimensionless functions. The scalar and gauge field equations of motion (EOM) are 
\beqa
\dfrac{d^{2}\phi}{d\rb^{2}} + \dfrac{2}{\rb}\frac{d\phi}{d\rb} &=& \dfrac{2\,u^{2}\,\phi}{\rb^{2}} + \frac{\lambda}{g^{2}}\phi\,(\phi^{2}-1) \; , \label{eq:phieqn} \\
\dfrac{d^{2}u}{d\rb^{2}} &=& \dfrac{u\,(u^{2}-1)}{\rb^{2}} + u\,\phi^{2} \; , \label{eq:gaugefieldeqn} 
\eeqa
where the dimensionless radius scale is defined as $\rb\equiv g\,f\, r = m_{W'}\, r $. Hence the only quantity of importance is the ratio $\lambda/g^{2}$. Boundary conditions are
\begin{equation}
\phi(0) = 0 \; , \quad  \phi(\infty) = 1 \; , \quad  u(0) =1 \; , \quad u(\infty) = 0 \; .
\end{equation}

Summing energies in both the gauge and scalar fields, the monopole mass is given by
\beqa
\label{eq:Mmono}
\Mmono &=& \int 4\,\pi\,r^2\left( \frac{1}{2}\,B^a_i \,B^a_i + \frac{1}{2}\, (D_i \Phi^a) (D_i \Phi^a) + V(\Phi) \right) \;,  \\
&=& \frac{4\pi f}{g} \int d\rb \rb^2 \left( \frac{\rb^2\,\phi'^{2} + 2\,u^{2}\phi^{2} }{2\,\rb^2}  + \frac{(1-u^{2})^{2} + 2\,\rb^2\,u'^{2}}{2\,\rb^{4}}  + \frac{\lambda}{4 g^{2}} (\phi^{2}-1)^{2}\right) \equiv \frac{4 \pi f}{g}\,Y(\lambda/g^2) \; ,\nonumber
\eeqa
where $Y$ is monotonic with $Y(0)=1$ (in the BPS limit~\cite{Bogomolny:1975de,Prasad:1975kr} with $\lambda \ll g^2$) and $Y(\infty) \approx 1.787$. Here, $'$ denotes a derivative with respect to $\bar{r}$. The magnetic field has the form 
\beqa
B^a_i = - \frac{1}{2}\,\epsilon^{ijk}\,F^a_{jk} \,\overset{r \rightarrow \infty}{\longeq}\, \hat{r}^a\,\hat{r}^i \, \frac{1}{g\,r^2} ~.
\eeqa
Since the $B$ field is along the $SU(2)$ space, one could choose the $\Phi^a$ VEV to be along a fixed direction, \eg, $a=3$, in the ``string gauge", which corresponds to the unbroken gauge direction or the dark photon direction. Then, the quantized magnetic charge
\beqa
Q_{\rm M} = 1\times g_{_\mathrm{M}} = \frac{4\pi}{g}  ~,
\eeqa
which satisfies the basic Dirac quantization condition with $Q_{\rm E}\,Q_{\rm M} = 2\pi$ for the smallest electric charge, $\frac{1}{2}g$, from a particle in the fundamental representation of $SU(2)$. 

Depending on the masses of $m_{W'}$ and $m_{h'}$, the gauge field and the scalar field have different radii to describe their profiles. At large $r$, the corresponding function $u$ for the gauge field behaves as $u \propto  e^{- m_{W'}\, r}$. The function for the scalar field at large $r$ goes as
\beqa
1 - \phi \simeq \frac{c_1}{m_{W'}\,r}\,e^{- m_{h'}\,r} \,+ \, \frac{c_2}{m_{h'}^2\,r^2}\,e^{- 2\,m_{W'}\,r } ~,
\eeqa
with $c_1$ and $c_2$ as two constants. Using the characteristic radius of the scalar field to label the size of the monopole, the monopole radius is
\beqa
\label{eq:radius-mono-one-scale}
\Rmono \simeq \mbox{min}[m_{h'}^{-1}, m_{W'}^{-1}]  ~.
\eeqa
For the case with $m_{h'} \ll m_{W'}$, the scalar and gauge profiles have a similar radius $\sim \mwd^{-1}$. In the other limit with $m_{h'} \gg m_{W'}$, the scalar profile has a much smaller radius than the gauge field. 

\subsection{Higgs-portal dark monopole}
\label{sec:simpleHiggs}
At the renormalizable level, the Higgs-portal coupling to the dark sector with a dark monopole is the only allowed one.~\footnote{For other topological defects like dark string, one can also have a kinetic mixing between dark $U(1)$ and SM hydercharge gauge fields~\cite{Hyde:2013fia}.} Therefore, we will treat it as the dominant coupling between the SM and dark sectors. The most general renormalizable potential for $\Phi$ and the SM Higgs field $H$ is
\beqa
\label{eq:VsimpleHiggs}
V(\Phi, H) = \frac{\lambda_\phi}{4} \, |\Phi|^4 - \frac{1}{2}\,\mu_\phi^2 \,|\Phi|^2 + \lambda_h \, (H^\dagger H)^2 + \mu_h^2\,H^\dagger H - \frac{1}{2}\, \lambda_{\phi h} |\Phi|^2  H^\dagger H + V_0 \; ,
\eeqa
where $V_0$ is a constant to make $V=0$ when the fields are at their global minima. Requiring the potential is bounded from below, one has the constraint $\lambda_h,\lambda_\phi>0$ and $2 \sqrt{\lambda_h \lambda_\phi} > \lambda_{\phi h}$. Since the SM Higgs boson mass $m_h \approx 125$~GeV provides us $\lambda_h \approx 0.13$, this condition is approximately $\lambda_\phi \gtrsim 1.9\,\lambda_{\phi h}^2$. Furthermore, requiring the vacuum to spontaneously break both the dark $SU(2)$ and the SM EW gauge symmetries,
\beqa
\label{eq:vev-condition}
- 2 \, \lambda_\phi \,\mu_h^2 + \lambda_{\phi h}\, \mu_\phi^2  > 0  \,,  \qquad \qquad
2 \, \lambda_h \, \mu_\phi^2  - \lambda_{\phi h} \, \mu_h^2  > 0 \,.
\eeqa
Defining $\langle |\Phi| \rangle = f$ and $\langle H^\dagger H \rangle= v^2/2$, one has the VEV's at the global vacuum as
$v^2 = (2\lambda_{\phi h}\mu_\phi^2 - 4 \lambda_\phi \mu_h^2)/(4\lambda_h \lambda_\phi - \lambda_{\phi h}^2)$  and $f^2 = (4 \lambda_h \mu_\phi^2 - 2 \lambda_{\phi h} \mu_h^2 )/(4 \lambda_h \lambda_\phi - \lambda_{\phi h}^2)$,  which can be rewritten as
\beqa
f^2 = \dfrac{2\,\mu_\phi^2 + \lambda_{\phi h} v^2}{2\,\lambda_\phi}\,, \qquad \qquad v^2=\dfrac{- 2\,\mu_h^2 + \lambda_{\phi h} f^2}{2\,\lambda_h} ~.
\label{eq:VEVs-one-scale}
\eeqa

The result of interest here is that in the interior of a dark monopole, $|\Phi|$ ranges from 0 to $f$, and so the Higgs vacuum value will also change inside of a monopole. Depending on the signs of various model parameters and subject to the constraints in \eqref{eq:vev-condition}, we have four different cases
\beqa
\mbox{Case I}: \qquad \mu_\phi^2 > 0\,, \qquad \mu_h^2 > 0 \,, \qquad \lambda_{\phi h} > 0 \,,  \nonumber \\
\mbox{Case II}: \qquad \mu_\phi^2 > 0\,, \qquad \mu_h^2 < 0 \,, \qquad \lambda_{\phi h} > 0  \,, \nonumber  \\
\mbox{Case III}: \qquad \mu_\phi^2 > 0\,, \qquad \mu_h^2 < 0 \,, \qquad \lambda_{\phi h} < 0  \,,  \nonumber  \\
\mbox{Case IV}: \qquad \mu_\phi^2 < 0\,, \qquad \mu_h^2 < 0 \,,  \qquad \lambda_{\phi h} > 0  \,.
\label{eq:4-cases}
\eeqa
Given the transition of the $\Phi$ field from 0 inside the monopole to $f$ outside, the EW VEV changes depending on the distance from the monopole core. For Case I, there exists a critical radial distance, $R_{\rm S}$, below which $v=0$ and the EW symmetry is restored. This is roughly given by $\lambda_{\phi h} f^2\, \phi(R_{\rm S})^2/2 = \mu_h^2$, with the determination of the precise location of $R_{\rm S}$ requiring a numerical calculation. For Case II, the EW symmetry is always broken throughout the monopole state and the EW VEV  increases from $- \mu_h^2 / \lambda_h$ at the core to $(\lambda_{\phi h} f^2/2 - \mu_h^2)/\lambda_h$  far away from the monopole. Similarly, for Case III, the EW VEV decreases from $- \mu_h^2 / \lambda_h$ at the core to $(\lambda_{\phi h} f^2/2 - \mu_h^2)/\lambda_h$ at the far away region. For Case IV, both symmetry breakings are triggered by the negative $\mu_h^2$. The behavior of the EW VEV is similar to Case II: monotonically increasing from core to outside. 

To obtain the quantitative results, we solve the coupled EOM for the dark gauge field, dark Higgs field, and SM Higgs field given by [using (\ref{eq:fieldRedef}) and rescaling $\sqrt{H^\dagger H} = h(r) v / \sqrt{2}$ to have a dimensionless profile function $h(r)$]
\beqa
\frac{d^{2}\phi}{d\rb^{2}} + \frac{2}{\rb}\frac{d\phi}{d\rb} &=& \frac{2\,u^{2} \phi}{\rb^{2}} + \frac{\lambda_{\phi}}{g^{2}}\phi^{3} - \frac{\mu_{\phi}^{2}\,\phi}{g^{2}f^{2}} - \frac{\lambda_{\phi h}}{2 g^{2}} \frac{v^{2}}{f^{2}}\,\phi\,h^{2} \;, \\
\frac{d^{2}u}{d\rb^{2}} &=& \frac{u(u^{2}-1)}{\rb^{2}} + u\,\phi^{2} \label{eq:ueqninefr} \;, \\
\frac{d^{2}h}{d\rb^{2}} + \frac{2}{\rb}\frac{dh}{d\rb} &=& \frac{\lambda_{h}}{g^{2}}\frac{v^{2}}{f^{2}} h^{3} + \frac{\mu_{h}^{2}}{g^{2}f^{2}}h - \frac{\lambda_{\phi h}}{2\,g^{2}} \phi^{2}\,h \;. \label{eq:eom-higgs}
\eeqa
The boundary conditions are 
\begin{equation}\label{eq:BCH}
\phi(0) = 0 \;,\; \phi(\infty) = 1 \;,\; h'(0) = 0\;,\; h(\infty) = 1\;, \; u(0)=1 \; , \; u(\infty) = 0 \; .
\end{equation}
Note the Neumann boundary condition for $h$.  This prevents the solution for $h$ from diverging as $\rb \rightarrow 0$.

Since the dark Higgs enters the SM Higgs EOM in \eqref{eq:eom-higgs}, the behavior of the SM Higgs profile naturally follows $\phi(r)$ and is determined by $\Rmono$ via \eqref{eq:radius-mono-one-scale}. Also, based on our stated phenomenological interests, we would like to explore the parameter space with different EW VEVs in a localized region with a large radius compared to the EW scale. This amounts to requiring $\Rmono  \gg v^{-1} \sim 10^{-2}\,\mbox{GeV}^{-1}$ or 
\beqa
m_{h'}\,, m_{W'} \ll v ~. 
\eeqa
Given that $m_{W'} = g f$ and $m_{h'} \approx\sqrt{2 \lambda_\phi} f$ (in the limit of small mixing between dark Higgs and SM Higgs), a large monopole radius can be achieved by $f < v$ and/or by small $\sqrt{\lambda_\phi}, g \ll 1$.  We will refer to these cases as ``small $f$'' and ``small $g$," respectively.  

\begin{figure}[tb!]
	\label{fig:one-scale-cases}
	\begin{center}
		\includegraphics[width=0.48\textwidth]{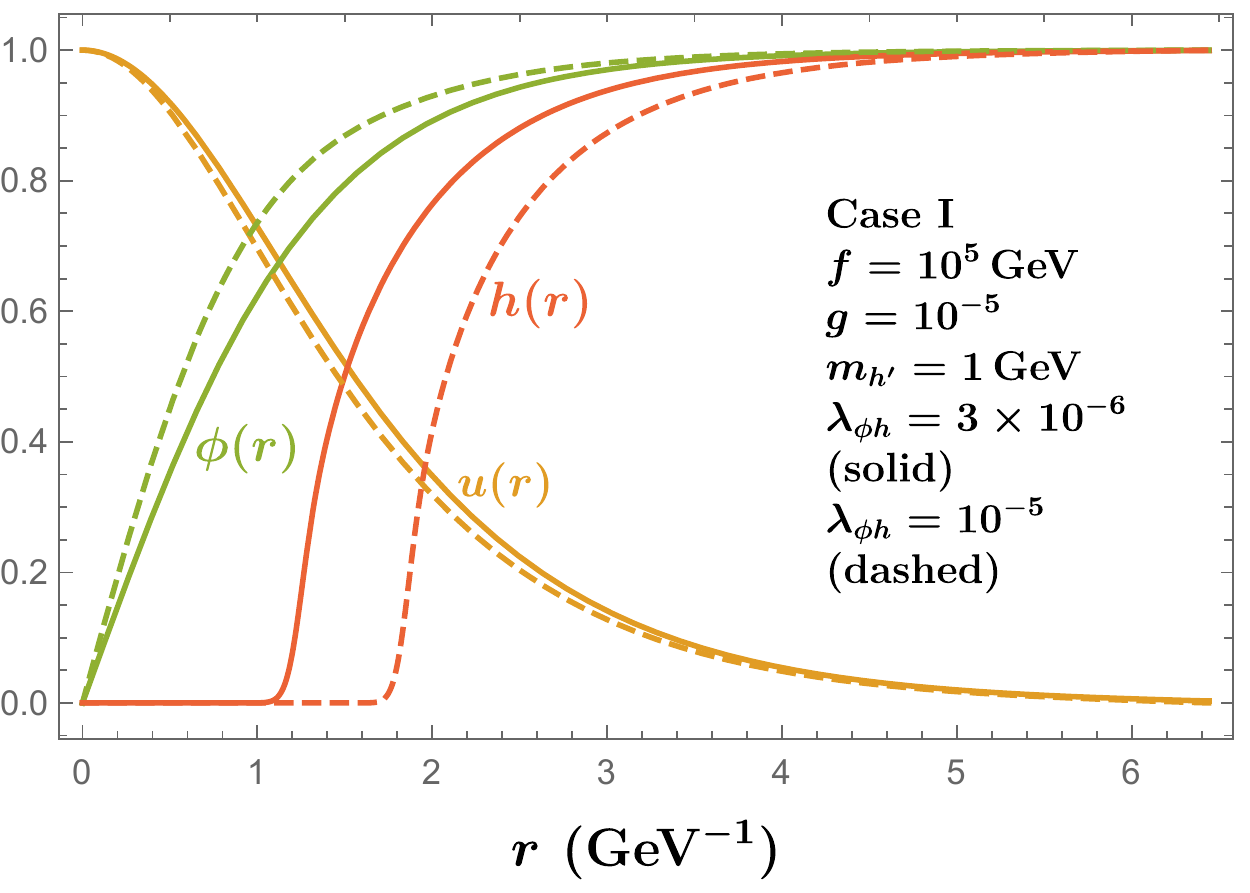} \hspace{3mm} 
		\includegraphics[width=0.48\textwidth]{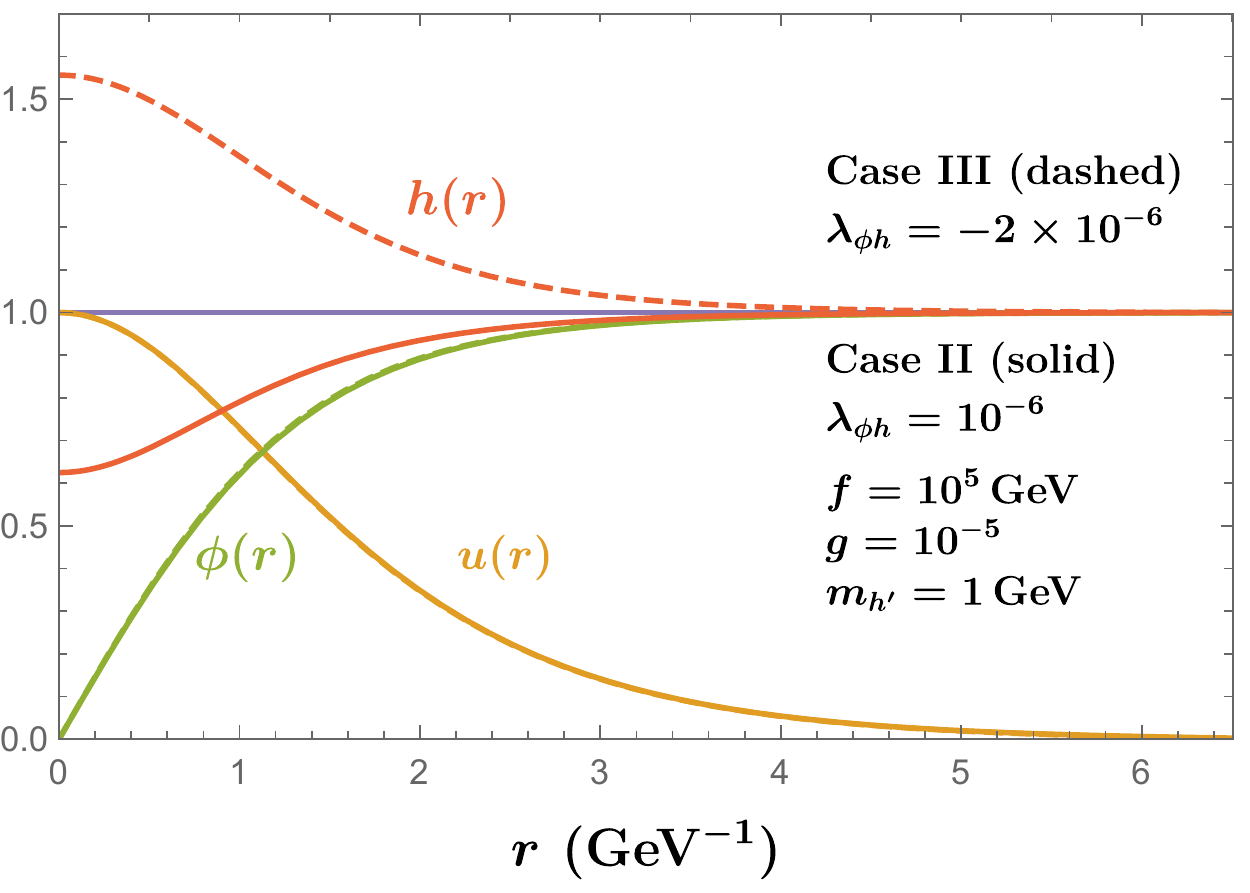}
		\caption{Representative profiles as a function of radius for various cases defined in \eqref{eq:4-cases} are shown here in the ``small $g$'' scenario. Parameters have been chosen to obtain $v=246~\GeV$ and $m_h=125~\GeV$ in all cases.}
	\end{center}
\end{figure}

In the small $g$ (and large $f \gg v$) scenario for Case I in \eqref{eq:4-cases}, the EW symmetry can be restored in the interior of the monopole even for small $\lambda_{\phi h} \ll 1$ required by decoupling of the dark sector (see Sec.~\ref{sec:preheating}).  Numerically, we show example field profiles in the left panel of Figure~\ref{fig:one-scale-cases} for Case I. One can see that the EW restoration region is larger for a larger portal coupling $\lambda_{\phi h}$. For generic parameters, the EW-restoration radius $R_{\rm S}$ matches the dark monopole radius $R_{\rm S} \sim \Rmono$. On the other hand, for a tuned parameter choice with $v^2 \ll \mu_h^2, \lambda_{\phi h} f^2$ and a delicate cancellation between $2 \mu_h^2$ and $\lambda_{\phi h} f^2$ via \eqref{eq:VEVs-one-scale} to obtain the EW scale, one could have $R_{\rm S} \gg \Rmono$. Also from this plot, one can see that the modifications to the monopole profiles when including the Higgs-portal coupling are small. Both $u(r)$ and $\phi(r)$ profiles do not change at the qualitative level for these numerical examples. In the right panel of Figure~\ref{fig:one-scale-cases}, representative profiles for the Case II and Case III are also shown, while the profiles for Case IV are similar to Case II. For these cases, the EW symmetry inside the monopole is not restored but broken less (Cases II,IV) or broken more (Case III). 

On the other hand, in the small $f$ scenario, the EW symmetry cannot be restored inside the monopole because $\lambda_{\phi h}$ would be impermissibly large and far beyond the perturbative region.  Thus, in this scenario $\mu_h^2 < 0$ is very nearly its SM value.  Inside the monopole, the Higgs field will differ only slightly from its vacuum expectation value.  On the other hand, all gauge and quartic couplings can be $\mathcal{O}(1)$, whereas they must be tiny in the small $g$ case. The small change to the Higgs field is difficult to detect using conventional search strategies.  Thus, we will focus on the first scenario with a small $g$ for the phenomenology prospects in Section~\ref{sec:pheno}.

\section{Monopoles  from the Kibble-Zurek mechanism}
\label{sec:kibble-zurek}

In this section, we review the Kibble-Zurek mechanism for production of topological defects during a thermal phase transition in the dark sector~\cite{Kibble:1976sj,Zurek:1985qw}. In this mechanism, only point-like dark monopoles with radii much smaller than the EW scale can account for 100\% of DM, while larger monopoles will have suppressed abundances. This is because the monopole-antimonopole annihilation rate in the thermal dark $W'^\pm$ plasma is enhanced for larger-radii monopoles.  

For the monopole abundance from the Kibble-Zurek mechanism~\cite{Kibble:1976sj,Zurek:1985qw}, monopoles are initially produced during a first or second order phase transition, starting at some critical temperature in the dark sector $T'_c$ (where primed denotes dark sector and unprimed denotes visible sector), often of order the symmetry-breaking scale $T'_c \simeq f$. Then, the monopole number density $n_{\tiny \Mcirc}$ is related to the correlation length $\xi$ of the symmetry-breaking field by $n_{\tiny \Mcirc} \sim  \xi^{-3}$. The correlation length is smaller than the Hubble size, $\xi < d_{H}(T'_c)$,  at the end of the nucleation process with $T'_n \simeq T'_c$. If one neglects the potential temperature difference between the dark sector and the SM sector, the upper bound on the correlation length sets a lower bound on the monopole number density $n_{\tiny \Mcirc}(T'_c) \gtrsim d_{H}(T'_c)^{-3}$, which is known as the Kibble limit~\cite{Kibble:1976sj}. 

Depending on the order of the phase transition, a much larger number density than the Kibble limit is anticipated.  For the first-order phase transition, the number of nucleation sites with one Hubble patch could be a larger number. The nucleated bubble radius provides a more precise estimation of the correlation length: $\xi \sim r_{\rm bubble} \simeq (M_{\rm pl}/T_c^{\prime 2})/\ln(M_{\rm pl}^4/T_c^{\prime 4}) \ll d_{H}(T'_c)$ for $T'_c \ll M_{\rm pl}$~\cite{Guth:1982pn}. The relic monopole abundance, or the number density over the entropy, is estimated to be~\cite{Kolb:1990vq}
\beqa
\label{eq:KibbleYield}
Y(T_c)\equiv \frac{n_{\tiny \Mcirc}}{s} \simeq g_{*s}^{-1}\, \kappa^3\,\left[ \left(\frac{T'_c}{M_{\rm pl}}\right)\, \ln \left( \frac{M_{\rm pl}^4}{T_c^{\prime 4}}\right) \right]^3 ~,
\eeqa
where we have used $s=(2\pi^2/45) \, g_{*s}\,T^3 \approx 0.44\,g_{*s}\,T^3$ and introduced a dimensionless parameter $\kappa \equiv T'/T$ to account for the possibility of a colder dark sector. 
For a second-order phase transition, the calculation was refined by Zurek~\cite{Zurek:1985qw} to account for the finite relaxation time of the system as $\xi$ formally diverges near $T_c$. The resulting yield is
\beqa
\label{eq:yieldKibbleZurek}
Y(T_c) &\simeq & g_{*s}^{-1}\,g_*^{\frac{3\nu}{2(1+\mu)}} \kappa^{\frac{3(1+\mu-2\nu)}{(1+\mu)}} \, \lambda^{\frac{3(1+\mu-\nu)}{2(1+\mu)}}\, \left( \frac{T_c^\prime }{M_{\rm pl}} \right)^{\frac{3\nu}{1+\mu}} = g_{*s}^{-1}\,g_*^{1/2}\,\kappa\,\lambda\,\frac{T_c^\prime }{M_{\rm pl}}~,
\eeqa
where $\mu$ and $\nu$ are the critical exponents for the relaxation time and the correlation length as a function of $(T_c - T)/T_c$~\cite{Murayama:2009nj}. In the last equality, the classical values of $\mu=\nu=1/2$ have been substituted, although quantum corrections modify these exponents \cite{Murayama:2009nj,Khoze:2014woa}. Notice that the yield is enhanced compared to (\ref{eq:KibbleYield}) by a factor $(\Mpl/T'_c)^{2}$, and there is an additional dependence on $\lambda$. If $\lambda$ and $\kappa$ are not too small and $T_c' \propto f$, a second order phase transition will admit much smaller values for $f$ to make up all of DM than a first order phase transition.

After being produced during a phase transition, monopoles and antimonopoles will partially annihilate, reducing their relic abundance today~\cite{Preskill:1979zi}.  Their annihilation rate is heavily influenced by the presence (or absence) of a (dark-)charged plasma.  Within a plasma, the monopole free-streaming length is reduced, resulting in an attractive drift velocity between monopole-antimonopole pairs that enables their annihilations.  Without a plasma, the monopoles free stream and rarely encounter each other. Thus, the abundance of the $W'^\pm$ is one of the crucial quantities to determine the final abundance of dark monopole. The monopole and antimonopole annihilations will effectively freeze out at temperature $T'_F$ when the charged $W'^\pm$ becomes non-relativistic and has a reduced abundance. As emphasized in Ref.~\cite{Preskill:1979zi}, the final monopole abundance in the current Universe could be independent of its initial abundance just after the phase transition. 

The motion of monopoles in plasma is characterized by two scales: the free streaming length, $l_\text{free}$, and the capture radius, $l_\text{capt}$. The free streaming length is proportional to the interaction length multiplied by the number of interactions that are necessary to change the monopole momentum direction
\beqa
\ell_{\rm free} \sim \frac{p_{\tiny \Mcirc} }{p_{W'}}\, \frac{1}{\sigma_{\rm scat}\,n_{W'}} ~.
\eeqa
For a thermalized dark sector, monopoles and $W'^{\pm}$ share the same temperature $T'$, which could be different from the SM sector. For small $g \ll 1$, the $W'^\pm$ mass is smaller than the phase transition temperature $T'_c$, so initially $W'^\pm$ behave as relativistic particles. One has $p_{\tiny \Mcirc} \simeq \sqrt{T' \, M_{\tiny \Mcirc} }$, $\sigma_{\rm scat} \simeq 4\pi/T'^2$, and $n_{W'} = (\zeta(3)/\pi^2) \gstarc\, T'^3$, with $\gstarc \geq 6$ the number of charged relativistic degrees of freedom (depending on the other field content of the dark sector). Defining a dimensionless quantity $B= \sigma_{\rm scat}\,n_{W'}/T' \approx 4 \zeta(3)\,\gstarc/\pi \approx 2.2$, the free streaming length is estimated to be 
\beqa
\ell_{\rm free}  \sim \frac{1}{B}\, \frac{M_{\tiny \Mcirc}^{1/2} }{T'^{3/2} } = \frac{\sqrt{4\pi} }{B}\, \frac{f^{1/2}}{g^{1/2}\,T'^{3/2} }  ~.
\eeqa
The capture length of one monopole by another antimonopole can be estimated to be the distance at which the potential energy equals the kinetic energy
\beqa
\ell_{\rm capt} \sim \frac{g_\mathrm{M}^2}{4\pi\,T'} = \frac{4\pi}{g^2\,T'} ~.
\eeqa
Just after the phase transition with $T' = T'_c \simeq f$, one has $\ell_{\rm free}/\ell_{\rm capt} \simeq g^{3/2}/(\sqrt{4\pi}) \ll 1$ for $g \ll 1$, so monopole-antimonopole bound states can form.  As the temperature drops, this ratio increases. Therefore, we can define a ``freeze-out"  temperature, $T'_{F}$, with $\ell_{\rm free} \simeq \ell_{\rm capt}$. This can occur in one of two ways.  First, if the number density of dark-charged particles is suppressed, $l_\text{free} \propto 1/n_{W'}$ naturally becomes large.  This may happen, \eg, when $W'^\pm$ become non-relativistic and their in-equilibrium number density is exponentially depleted.  If this is realized, $T'_F$ should be of order the $W'^\pm$ mass or $T'_F \simeq g f$. Second, while $W'^\pm$ remains relativistic until $T'_F$, the equality of the two length scales, $\ell_{\rm free}\simeq \ell_{\rm capt}$,  determines $T'_F \simeq g^3\,f/(4\pi B^2)$. For $g \lesssim 1$, the freeze-out temperature is determined by the first step of $T'_F \simeq g f$.

During the time with $\ell_{\rm free} <  \ell_{\rm capt}$, the monopole-antimonopole bound state formation rate is related to their relative drift velocity. The drag force is proportional to the drift velocity $v_{\rm drift}$ as $F_{\rm drag} \sim \sigma_{\rm scat}\, n_{W'}\, T'\,v_{\rm drift} = B\,T'^2\,v_{\rm drift}$. Equalizing the drag force and the attractive force, $v_{\rm drift} \sim g_\mathrm{M}^2/(4\pi\,B\,T'^2\,n_{\tiny \Mcirc}^{-2/3})$, where we have used the typical monopole separation distance $n_{\tiny \Mcirc}^{-1/3}$. The annihilation rate is then estimated to be
\beqa
\Gamma_{\rm drift} \simeq \frac{v_{\rm drift} }{n_{\tiny \Mcirc}^{-1/3} } \sim \frac{4\,\pi\,n_{\tiny \Mcirc}}{g^2\,B\,T'^2} ~. 
\eeqa
We can write the Boltzmann equation for the monopole number density
\beqa
\dot{n}_{\tiny \Mcirc} + 3\, H\, n_{\tiny \Mcirc} \approx  - \Gamma_{\rm drift} \, n_{\tiny \Mcirc} \sim - \frac{4\pi}{g^2\,B\,T'^2}\, n_{\tiny \Mcirc}^2 ~.
\eeqa
Solving this equation, one has the yield as a function of $T$ as 
\beqa
Y(T)^{-1} = Y(T_c)^{-1} + \frac{4\pi\,g_{*s}\,M_{\rm pl} }{\sqrt{g_*}\,B\,\kappa^2\,g^2} \left(T^{-1} - T_c^{-1}\right) ~,
\eeqa
Here, we have $\kappa = T'/T$ to denote the possible temperature difference between the dark sector and SM sector. The annihilation process approximately stops at the freeze-out temperature $T'_F$ or $T_F = T'_F/\kappa$. With $T'_F \ll T'_c$ (or equivalently $T_F \ll T_c$) and assuming an initial larger abundance, the freeze-out yield has a simple formula:
\beqa
\label{eq:yieldAnn}
Y(T_F) \approx  \frac{\sqrt{g_*}\,B\,\kappa^2\,g^2}{4\pi\,g_{*s}\,M_{\rm pl} } T_F ~.
\eeqa
This result could be approximately obtained by setting $\Gamma_\text{drift} = H$ at $T=T_F$. Fitting to the observed dark matter relic abundance, $\Omega_{\rm dm} h^2 \approx 0.120$~\cite{Aghanim:2018eyx}, the required yield is $Y(T_F) \approx 3.6\times 10^{-10}\,(1\,\mbox{GeV}/m_{\rm dm})$. Applying this result to the dark monopole case, we have 
\beqa
\Omega_{\tiny \Mcirc}h^2 &\approx & 0.120\times \frac{\sqrt{g_*}\,B\,\kappa}{4\pi\,g_{*s}\,\times 3.6\times 10^{-10}} \, \frac{g^2\, T'_F\,\Mmono}{M_{\rm pl}\times \mbox{1\,GeV}} \nonumber \\
&\approx & 0.112\,\times\,\left( \frac{\kappa}{1/10}\right)\, \left(\frac{g\,f}{1.5\times 10^{6}\,\mbox{GeV}} \right)^2 ~.
\label{eq:abundanceAnn}
\eeqa
In the second line of this equation, we have chosen $g_* \approx g_{*s}\approx 100$. 

One can see that to have dark monopoles account for 100\% of dark matter based on the Kibble-Zurek mechanism, the dark monopole radius is approximately bounded from above $\Rmono \approx (g \, f)^{-1} \lesssim  7 \times 10^{-7}\,\mbox{GeV}^{-1}$, which is shorter than the weak scale.

\section{Monopoles from preheating}
\label{sec:preheating}

In this section, we study a non-thermal production mechanism for dark monopoles. More specifically, we will adopt the parametric resonance mechanism during the inflaton preheating stage of our Universe~\cite{Dolgov:1989us,Traschen:1990sw,Kofman:1994rk,Kofman:1997yn}.  We will assume that the non-Abelian symmetry is broken before the end of inflation such that there is already a nonzero VEV $\langle |\Phi| \rangle = f$. Because of inflation, the symmetry-breaking direction of $\Phi$ is uniform within our Hubble patch so that no topological defects can form.
After inflation, the coupling of $\Phi$ to the oscillating inflaton field, $\mathcal{I}$, grows the fluctuations of $\Phi$ until all possible symmetry-breaking phases are populated in different subhorizon patches of the Universe.
As a result, the necessary topological configuration for dark monopoles is formed (see earlier studies in Refs.~\cite{Kasuya:1997ha,Khlebnikov:1998sz,Kasuya:1998td,Tkachev:1998dc,Kasuya:1999hy,Rajantie:2000fd,Kawasaki:2013iha} for production of other topological defects like cosmic strings and domain walls via preheating). 
The main difference from the previous Kibble-Zurek mechanism is that the  phase transition happens when the dark sector energy density scale is dramatically smaller than the symmetry breaking scale $f$, such that the dark monopole free-streaming length is increased relatively compared to the capture length onto an antimonopole. Dark monopoles produced in this mechanism can easily accommodate 100\% dark matter abundance.

Via parametric resonance, we require that only a small fraction of the inflaton energy is transferred into the dark sector. The majority of the energy will be transferred into our SM sector at a later time via either the parametric resonance or inflaton-decay reheating mechanism. For simplicity, we introduce an inflaton decay width of $\Gamma_{\mathcal{I}}$, so the reheating temperature in the SM sector is $T_\mathrm{RH} \sim \sqrt{\Gamma_{\mathcal{I}}\,M_\mathrm{pl}}$~\cite{Bassett:2005xm}. We will consider the separate cases where the dark sector is internally non-thermalized or thermalized. Similar to the Kibble-Zurek case, we will assume that the interactions between the dark sector and SM sector are weak enough such that the two sectors are not thermalized with each other.

%
\subsection{Parametric resonance}
\label{sec:parametric-resonance}

In this subsection, we apply the general parametric resonance mechanism to our dark monopole model. For our purposes, the details of the inflaton potential $V(\mathcal{I})$ during inflation are unimportant. 
We will leave the details of inflation unspecified and instead concentrate on the later stage of inflaton dynamics when it starts to oscillate around the minimum of its potential, which we take as quadratic for simplicity. Further from the origin, it must deviate from quadratic if it acts as the inflaton in order to satisfy constraints on the tensor-to-scalar ratio \cite{Akrami:2018odb}.  Note that parametric resonance may start well after the end of inflation.

The coupling between the inflaton field $\mathcal{I}$ and the $\Phi$ field is described by the potential terms
\beqa
 V \supset \frac{1}{2}\,m_{\mathcal{I},0}^2 \,\mathcal{I}^2\,+\,\frac{1}{2}\,\lambda_{\mathcal{I}\phi}\,\mathcal{I}^2\,|\Phi|^2 ~,
\eeqa
where the interaction operator conserves a $\mathcal{Z}_2$ symmetry of $\mathcal{I} \rightarrow - \mathcal{I}$. Here, $m_{\mathcal{I},0}$ is the bare inflaton mass. The total inflaton mass after the breaking of $SU(2)$ (when $\Phi$ obtains its VEV $f$) is $m_\mathcal{I}^2 = m_{\mathcal{I},0}^2 + \lambda_{\mathcal{I}\phi} f^2$. In our study later, we will assume $\lambda_{\mathcal{I}\phi} f^2 \gg m_{\mathcal{I},0}^2$. The inflaton potential is assumed to be locally quadratic near the origin. When the inflaton oscillates, it behaves as a matter field, with its field value approximately described by $\mathcal{I}= \mathcal{I}_0 (a_0/a)^{3/2}\cos[m_\mathcal{I}\,(t - t_0)]$, with $t_0 \sim m_\mathcal{I}^{-1}$ as the starting time of oscillation, $a(t=t_0)=a_0\equiv 1$ as the initial scale factor, and $\mathcal{I}_0$ as the initial inflaton amplitude. Requiring the $SU(2)$ symmetry is broken at $t=t_0$, the inflaton amplitude needs to satisfy $\lambda_{\mathcal{I}\phi}\, \mathcal{I}_0^2 < \lambda f^2$, where $\lambda$ is the $\Phi$ self-quartic coupling defined in \eqref{eq:basic-dark-lag}. In the monopole abundance calculation, we ignore the $\Phi$ interactions with the SM Higgs, which do not change the parametric resonance discussion here. 

At the classical level, the $SU(2)$ spontaneous symmetry breaking has a homogenous phase throughout the Universe. No non-trivial topological field configuration exists in the Universe. At the quantum level, on the other hand, small but inhomogeneous primordial quantum fluctuations exist with perturbed amplitude $\simeq H_\mathcal{I} /(2\pi)$, with $H_\mathcal{I}$ the Hubble parameter during inflation. These primordial fluctuations can grow exponentially because of the oscillating inflaton field. To study the early growth of modes, one can linearize the $\Phi_a$ and $\mathcal{I}$ EOM. Without loss of generality, we assume the initial symmetry-breaking direction is along the $\Phi_1$ direction with $\langle \Phi_1\rangle = f$. The perturbed field around the vacuum is then defined as $\delta \Phi_1 = \Phi_1 - f$, $\delta \Phi_2 = \Phi_2$, $\delta \Phi_3 = \Phi_3$. In the limit of $\lambda_{\mathcal{I} \phi}\, \mathcal{I}_0^2 \ll \lambda f^2$, one can ignore the background modifications from the inflaton field. The EOM of $\delta \Phi_i$ in Fourier space at linear order in the perturbations are
\beqa
\label{eq:delta-Phi_1}
\ddot{\delta \Phi_1}_k + 3 H \,\dot{\delta \Phi_1}_k + \frac{k^2}{a^2} \,\delta \Phi_{1k} +  ( 2 \lambda \,f^2 + \lambda_{\mathcal{I}\phi}\,\chi^2 )\, \delta \Phi_{1k}  &=& 0 \, , \\
\label{eq:delta-Phi_23}
\ddot{\delta \Phi_{2}}_{,3\,k} + 3 H \,\dot{\delta \Phi_{2}}_{,3\,k} + \frac{k^2}{a^2}\, \delta \Phi_{2,3\,k} + \lambda_{\mathcal{I}\phi}\, \chi^2\, \delta \Phi_{2,3\,k} &=& 0 \, , 
\eeqa
where we have used the co-moving momentum $k \equiv a\,p$ with $p$ as the ordinary momentum. Here, we have neglected terms related to the gauge couplings \cite{Rajantie:2000fd,GarciaBellido:2003wd}, as we expect the $\Phi$ dynamics to give the most important effect.

\begin{figure}[tb!]
	\label{fig:mathieu-diagram}
	\begin{center}
		\includegraphics[width=0.47\textwidth]{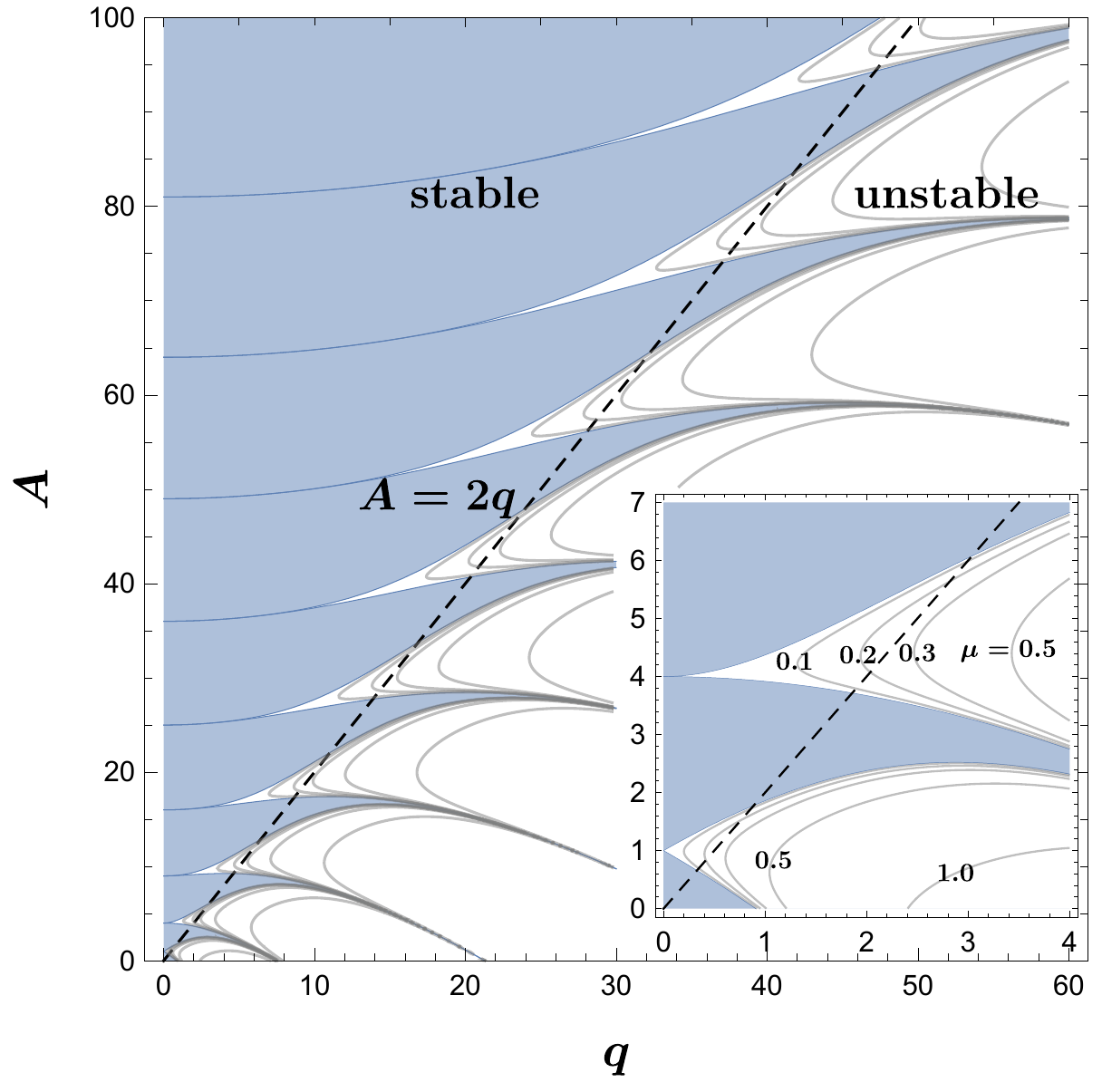} \hspace{3mm}
	\includegraphics[width=0.48\textwidth]{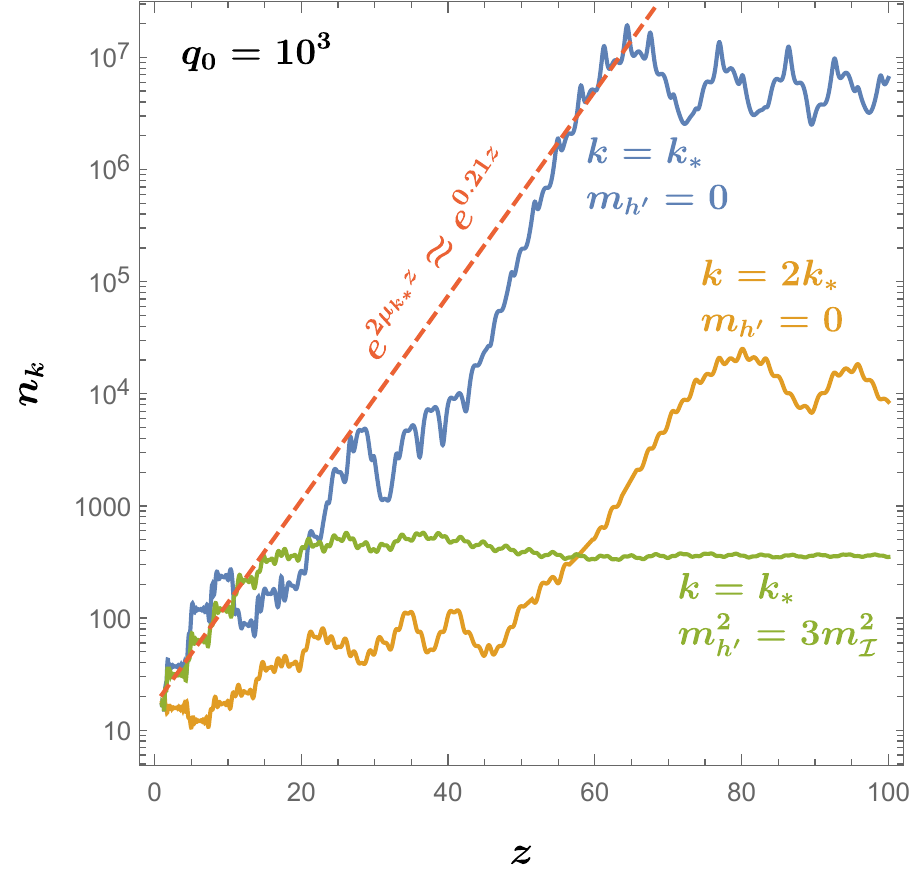}  
		\caption{{\it Left}: The stability diagram for the Mathieu equation in the plane of $q$ and $A$. For the unstable region, exponential growth solutions are expected with the exponential index $\mu=0.1, 0.2, 0.3, 0.5, 1.0$ from left to right shown in the gray contour lines. {\it Right}: Solution to the linearized equation of motion for $\delta \varphi_1$ expressed as the growth of the comoving number density $n_k$, showing the effect of varying $k$ in terms of $k_* \simeq m_\mathcal{I}\,q_0^{1/4}$ and the ratio $m_{h'}^2 / m_\mathcal{I}^2$ ($m_{h'}^2=0$ corresponds to the solution for $\delta \varphi_{2,3}$).}
	\end{center}
\end{figure}

To match to the Mathieu equations~\cite{Mathieu1868}, we define rescaled fields $\delta \varphi_{i, k} \equiv a^{3/2}\, \delta \Phi_{i, k}$ and a dimensionless time $z\equiv m_{\mathcal{I}} \,(t - t_0)$. Neglecting terms with $H^2 \sim \ddot{a}/a \ll m_{\mathcal{I}}^2$, \eqref{eq:delta-Phi_1} and \eqref{eq:delta-Phi_23} become
\beqa
\delta \varphi_{1\,k}'' \,+ \,\big[A_{1\, k} + 2\,q \,\cos(2z) \big]\, \delta \varphi_{1\,k} &=& 0  ~, \\
\delta \varphi_{2,3\,k}''\, +\, \big[ A_{2,3\, k} + 2\,q \,\cos(2z) \big] \, \delta \varphi_{2,3\,k} &=& 0  ~,
\eeqa
where $'$ denotes a derivative with respect to $z$. Here, the factors are defined as
\beqa
\label{eq:q-A-relation}
q_0 \equiv \frac{\lambda_{\mathcal{I}\phi} \, \mathcal{I}_0^2 }{4\,m_{\mathcal{I}}^2} ~, \quad  q \equiv \frac{q_0}{a^3} ~,\quad A_{1\,k} = 2\,q + \frac{k^2}{a^2\,m_{\mathcal{I}}^2} + \frac{m_{h'}^2}{m_{\mathcal{I}}^2} ~, \quad A_{2,3\,k} = 2\,q +\frac{k^2}{a^2\,m_{\mathcal{I}}^2} ~.
\eeqa
The stable and unstable (exponential growth) regions of the Mathieu equations for general $A$ and $q$ have been studied extensively, and they are plotted in the Strutt-Ince diagram~\cite{ince_1927}. For presentation purposes, we calculate this diagram numerically for general $q$ and $A$ and show it in the left panel of Fig.~\ref{fig:mathieu-diagram}. The region to have parametric resonance corresponds to the ``unstable" region with exponentially growing solutions (see Ref.~\cite{Bateman:100233} for analytic formulas).  Eq.~(\ref{eq:q-A-relation}) enforces $A \geq 2q$ for all modes and fields. For a very small $q \ll 1$, the unstable region of $A$ or $k$ is very narrow, which can be seen from the zoomed-in inset plot. Note that the $n$'th unstable band intersects the $A$-axis at $n^2$. For a larger $q \gtrsim 1$, a broader region of $A$ or $k$ can have parametric resonance. For a fixed $q$, only a few broad resonance bands with $A \geq 2q$ are anticipated. 

Because $A_{1\,k} \geq A_{2,3\,k}$ in \eqref{eq:q-A-relation}, it is much easier for $\delta\varphi_{2,3\,k}$ to sit in a broad resonance band than $\delta\varphi_{1\,k}$. To demonstrate this difference and also to calculate the exponential growth index, we compute the solution for $\delta \varphi_{1\,k}$ for different values of $m_{h'}^2/ m_\mathcal{I}^2$ and $k$ keeping the full scale factor dependence.  In the right panel of Fig.~\ref{fig:mathieu-diagram}, we display the comoving number density of particles with comoving momentum $k$ that are produced, given by \cite{Kofman:1997yn},
\beqa
\label{eq:PRnk}
n_k = \frac{\omega_k}{2} \left( \frac{|\dot{\delta \varphi}_k|^2}{\omega_k^2} + |\delta \varphi_k|^2 \right) - \frac{1}{2} ~,
\eeqa
with $\omega_k^2 = A_k + 2 q \cos (2 z)$.
When $m_{h'}^2 / m_\mathcal{I}^2=0$, the solution for $\delta \varphi_{1\,k}$ is equivalent to $\delta \varphi_{2,3\,k}$. At smaller $z$, the field fluctuations jump upwards and downwards stochastically as they enter resonance bands in/out of phase with the resonance, though they tend to increase over many such band-crossings.  This is the broad resonance regime when $q \gg 1$ as studied in \cite{Kofman:1997yn}.  Then, as $q$ becomes $\mathcal{O}(1)$, the massless fields enter a narrow-resonance regime where growth is strictly upwards and exponential (for the blue curve, this is between $40 \lesssim z \lesssim 60$).  The resonance ends when $q \lesssim 1/3$ leaves the lowest resonance band. In the $q_0 \gg 1$ regime demonstrated in the right panel of Fig.~\ref{fig:mathieu-diagram}, the strongest resonance has $k_* \simeq m_{\mathcal{I}}\, q_0^{1/4}$ with a width of $\Delta k_* \simeq k_*/2$~\cite{Kofman:1997yn}. When $k\gtrsim 1.5 k_*$, the final grown amplitude becomes smaller. The growth of field fluctuations for $k$ near to $k_*$ in the strongest resonant band can be approximated as $\delta \varphi_{2,3\,k_*} \propto e^{\mu_{k_*} z}$ with $\mu_{k_*} \simeq 0.13$ a typical value for a broad resonance (note, however, that $\mu_{k_*}$ is itself a stocastic quanitity, which is why the value in the right panel of Fig.~\ref{fig:mathieu-diagram} of $n_{k_*} \propto \delta \varphi_{k_*}^2 \propto e^{0.21 z}$ differs slightly). On the other hand, when $m_{h'} \gtrsim m_\mathcal{I}$, the growth of fluctuations is significantly suppressed compared to the massless case. Thus, at leading order we need only talk about fluctuations in $\Phi_{2,3}$ (approximately the Goldstone boson directions).

The field fluctuations in the coordinate space are estimated to be 
\beqa
\label{eq:fluctuation-amplitude}
\left<|\delta \Phi|^2\right> &\simeq & 2 \left<|\delta \Phi_{2,3}|^2\right> \simeq  \frac{1}{2\,a^3 \sqrt{\pi\,\mu_{k_*} z}} \left(\frac{H_\mathcal{I}}{2\pi}\right)^2 e^{2\,\mu_{k_*}z} ~.
\eeqa
where the initial fluctuations are assumed to be flat and determined by the quantum fluctuations induced during inflation: $\left<|\delta \Phi_k|^2\right>_0 \simeq 2\pi^2k^{-3}(H_\mathcal{I}/2\pi)^2$ with $H_\mathcal{I}$ the Hubble parameter during inflation.  The method of steepest descent was used to evaluate the Fourier transformation.
For $\Phi_{1}$, the fluctuations due to the parametric resonance are suppressed. One reason, $A_{1\,k} > A_{2,3\,k}$, has already been mentioned. The other reason is that its initial quantum fluctuations are also dramatically suppressed because the radial field mass must be greater than $H_\mathcal{I}$~\cite{Langlois:2004px}. However, after the amplitude of $\delta \Phi_{2, 3}$ grows to order $f$, the nonlinear terms in the $\delta \Phi_1$ EOM, {\it i.e.}, $\lambda\,f\,\delta \Phi_1\, (\delta \Phi_2^2 + \delta \Phi_3^2)$,  serve as a driving force to grow the fluctuations of $\delta \Phi_1$. Eventually, the ``phase" or actual location of the fields along the flat directions of the potential can have an order-one random value. Therefore, we will adopt the fluctuation amplitude in \eqref{eq:fluctuation-amplitude} and define the end of parametric resonance when $\left<|\delta \Phi|^2\right>\simeq f^2$. The curvature in the radial direction suppresses further growth, preventing $\left<|\delta \Phi|^2\right> \gg f^2$.  Up to this value, backreactions and rescatterings can be approximately neglected.  However, as we have just indicated, terms that are higher than linear order in the field fluctuations are ultimately important even for the symmetry restoration process, which introduces rescatterings \cite{Prokopec:1996rr}.  A full accounting requires detailed lattice simulations, beyond the scope of this work.  See Refs.~\cite{Kasuya:1997ha,Khlebnikov:1998sz,Kasuya:1998td,Tkachev:1998dc,Kasuya:1999hy,Rajantie:2000fd,Kawasaki:2013iha} for previous lattice work showing the validity of our approximations. See also Ref.~\cite{Vachaspati:2016abz} for an example of evolution of field configuration into monopoles.

Preheating ends when it leaves the lowest resonance band around $q \lesssim 1/3$. Thus, using that $a \simeq (t/t_0)^{2/3} \simeq z^{2/3}$ during matter domination, parametric resonance ends at $z_\text{end} \simeq (3 q_0)^{1/2}$ and $a_{\rm end} \simeq z_{\rm end}^{2/3}$.  To have order-one phase fluctuations, or $\left<|\delta \Phi|^2\right>\simeq f^2$, one has via \eqref{eq:fluctuation-amplitude} that
\beqa
\label{eq:q0min}
q_0 \gtrsim \frac{1}{3\,\mu_{k_*}^2}\,\left[ \log\left(q_0^{5/8}\,\frac{2\pi f}{H_\mathcal{I}}\right) \right]^2 ~,
\eeqa
which sets a lower bound on the value of $q_0 \approx \mathcal{I}_0^2/(4 f^2)$ from \eqref{eq:q-A-relation} when $m_\mathcal{I}^2 \approx \lambda_{\mathcal{I}\phi} f^2$. 

At the end of parametric resonance, the scalar and gauge field configurations can be different in regions separated by the correlation length. For the parametric resonance mechanism, the correlation length can be estimated as the typical wave-number with a large occupation number at the end of parametric resonance, or $ \xi \simeq p^{-1}_{*\,\mathrm{end}} = a_\mathrm{end}/k_*$. The dark monopole number density is related to the correlation length of the $\delta \Phi$ field fluctuations $\xi$ analogously to the Kibble mechanism, giving
\beqa
\label{eq:number-density-tend}
n_{\tiny \Mcirc}(t_\text{end}) \simeq \xi^{-3} \simeq p^3_{*\,\mathrm{end}} = (k_*/ a_\text{end})^{3} ~,
\eeqa
with $k_* \simeq m_\mathcal{I}\,q_0^{1/4}$. 

To determine the monopole abundance, we need to know how the SM sector is reheated or preheated by the inflaton field. To simplify our discussion, we will choose a simple inflaton-decay reheating scenario for the visible SM sector.~\footnote{Preheating to fermions is expected to be inefficient~\cite{Greene:2000ew}, and preheating to the Higgs is suppressed by the requirement that $m_\text{eff}$ be smaller than the dark charged gauge boson masses, which we  want to be lighter than the SM Higgs boson so that $\Rmono > 1/m_h$.}  
In order to not modify the preheating production of dark monopoles described before, the reheating to the SM must happen after the end of dark sector preheating. For this to be self-consistent, the preheating to the dark sector must finish before too much energy is transferred to the dark sector. 
\begin{figure}[tb!]
	\label{fig:timeline}
	\begin{center}
		\includegraphics[width=0.95\textwidth,height=3.05cm]{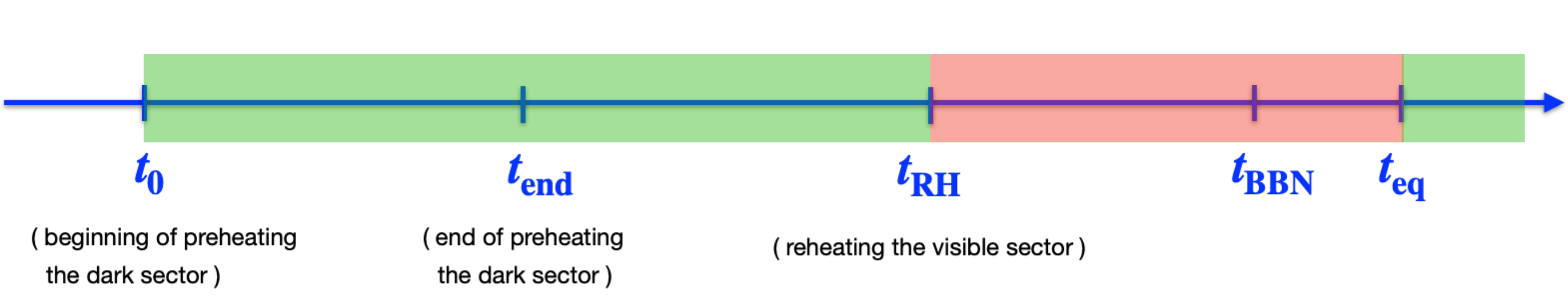}
		\caption{Schematic timeline (not to scale) for the inflaton preheating the dark sector and reheating the visible sector. From $t_0$ to $t_{\rm RH}$, the Universe has an early period of matter dominance with $H\propto a^{-3/2}$. From $t_{\rm RH}$ to the matter-radiation equality time $t_{\rm eq}$, the Universe is radiation dominated with $H\propto a^{-2}$. The matter (radiation) dominant period is labeled in green (red) color.}
	\end{center}
\end{figure}
In Fig.~\ref{fig:timeline}, we show the schematic timeline of the early Universe history in our setup. At $t_0$, inflation ends, the inflaton field $\mathcal{I}$ starts oscillating, and parametric resonance starts in the dark sector only; at $t_{\rm end}$, parametric resonance ends, the dark sector temperature (if thermalized) is smaller than the dark charged gauge boson masses; at $t_{\rm RH}$, $\mathcal{I}$ completes its decays to SM particles and reheating concludes at temperature $T_\text{RH}$ higher than the temperatures of Big Bang nucleosynthesis (BBN) $T_\text{BBN} \simeq \text{MeV}$ and matter-radiation equality $T_\text{eq} \simeq \text{eV}$. From $t_0$ to $t_{\rm RH}$, the Universe has an early period of matter dominance. After $t_{\rm RH}$, the Universe is radiation dominated until matter-radiation equality.

To calculate the monopole relic abundance in the current Universe, we first estimate its number density at $t_{\rm RH}$.  Between $t_\text{end}$ and $t_\text{RH}$ , the monopole yield $Y=n_{\tiny \Mcirc}/s$ is diluted by the inflaton dumping additional entropy into the SM during its decay \cite{Chung:1998zb}.  During reheating, the number density redshifts as $n_{\tiny \Mcirc} \propto a^{-3} \propto T^8$. At $t_{\rm RH}$, the radiation energy density approximately equals the inflaton energy density or $\rho_{\rm R}(t_{\rm RH}) \simeq \rho_{\mathcal{I}}(t_{\rm RH}) = \rho_{\mathcal{I}}(t_{\rm end})(a_\text{end}/a_\text{RH})^3$. At $t_{\rm end}$ with $q_{\rm end} \simeq 1/3$ and $\mathcal{I}_{\rm end} \simeq m_\mathcal{I}/\sqrt{\lambda_{\mathcal{I}\phi}} \simeq f$, one has $\rho_\mathcal{I}(t_{\rm end}) \simeq m^2_\mathcal{I}\,\mathcal{I}_{\rm end}^2/2$. The ratio of the scale factors is then,
\begin{equation}
\label{eq:aend-on-aRH}
\left(\frac{a_{\rm end}}{a_{\rm RH}}\right)^3 \simeq \frac{\pi^2\,g_{*\mathrm{RH}} T_{\rm RH}^4}{15\,\lambda_{\mathcal{I} \phi}\,f^4}
<1 \, ,
\end{equation}
which is less than one by the assumption $t_\text{end} < t_\text{RH}$ (see Fig.~\ref{fig:timeline}).  
Here, $g_{*\mathrm{RH}}$ is the radiation degrees of freedom at $t_{\rm RH}$. Together with the number density at $t_{\rm end}$ from \eqref{eq:number-density-tend}, one obtains the yield of dark monopoles at $t_{\rm RH}$ as 
\beqa
\label{eq:yield-preheat}
Y(T_{\rm RH}) = \frac{n_{\tiny \Mcirc}(T_\text{RH})}{s(T_\text{RH})} \simeq  \frac{1}{s(T_\text{RH})} \left(\frac{k_*}{a_\text{end}}\right)^3 \left(\frac{a_\text{end}}{a_\text{RH}}\right)^3 \simeq \lambda_{\mathcal{I}\phi}^{1/2}\, q_0^{-1/4}\,\left(\frac{T_\text{RH}}{f}\right) ~,
\eeqa
where the factor of $g_{*\mathrm{RH}}$ is approximately cancelled by the degrees of freedom in $s(T_\text{RH})$. 

If the monopole annihilations are suppressed or can be neglected (which will be justified later), the relic abundance of dark monopoles is then 
\beqa
\label{eq:abundance-preheat}
\Omega_{\tiny \Mcirc}h^2 & \simeq & 0.120\times \left(\frac{ \lambda_{\mathcal{I}\phi}^{1/2}\, q_0^{-1/4} }{3.6\times 10^{-10}}\right)\, \left(\frac{T_{\rm RH}\,\Mmono}{f\times 1\,\mbox{GeV}} \right) \nonumber \\
& \approx & 0.120\times \left( \frac{10^3}{q_0} \right)^{1/4}\,\left( \frac{\lambda_{\mathcal{I}\phi}^{1/2}/g}{2\times 10^{-7}} \right)\,\left(\frac{T_{\rm RH}}{1\,\mbox{MeV}}\right) \,,
\eeqa
which is independent of $f$ and different from the parameter dependence from the Kibble-Zurek mechanism. So, for very weak interaction coupling with $g \ll 1$, the dark monopole could have macroscopic radius and mass if $\lambda_{\cI\phi}$ and $T_\text{RH}$ are also small.  

%
\subsection{Constraints on model parameter space}
\label{sec:constraints}

To obtain the dark monopole abundance in \eqref{eq:abundance-preheat}, a few assumptions have been made. To be consistent, some model parameter space will be constrained to fulfill these assumptions. In this subsection, we first discuss some general requirements and then separate our discussion into two parts: a dark sector that is thermalized or non-thermalized with itself. The dark sector will always be assumed decoupled from the SM sector, which is controlled independently by $\lambda_{\phi h}$ in our model. In our analysis, we will take the $\Phi$ self-interacting quartic coupling $\lambda \simeq g^2$ and roughly treat $\lambda$ and $g$ as one model parameter, as would be the case in models of supersymmetry where $\lambda$ is generated by $D$-terms \cite{Martin:1997ns}.  This also has the effect of making $m_{h'} \simeq \mwd$, so both the $h(r)$ and $u(r)$ profiles have comparable radii and both terms in (\ref{eq:radius-mono-one-scale}) are comparable.  Note this choice is stable against radiative corrections, which would give $\lambda \sim g^4/16\pi^2$.

The first condition comes from the assumption that the dark $SU(2)$ symmetry is broken before or during inflation or $\lambda f^2 > \lambda_{\mathcal{I}\phi}\,\mathcal{I}_0^2$, which can be translated into 
\beqa
\label{eq:broken-before-inflation}
\frac{\lambda_{\mathcal{I}\phi}}{g^2}  < \frac{1}{4\,q_0} ~.
\eeqa

The second condition is to require the dark sector energy density to be a sub-dominant part of the total inflaton energy density while still having a long enough period of parametric resonance to produce monopoles. At $t_{\rm end}$ and using the energy density of quanta produced in the dark sector from Eqs.~(\ref{eq:PRnk}) and (\ref{eq:fluctuation-amplitude}), the total dark sector energy density is estimated to be~\cite{Kofman:1997yn}
\beqa
\rho_{\rm d}(t_{\rm end}) \simeq \frac{k_*^4}{8\,a_{\rm end}} \left(\frac{f}{H_{\cI}}\right)^2  \simeq  \frac{m_\mathcal{I}^4 \, q_0^{2/3}}{12}\, \left( \frac{f}{H_{\cI}}\right)^2 ~.
\eeqa
The energy density contained in the inflaton is $\rho_\mathcal{I}(t_{\rm end}) \simeq \frac{1}{2}\,m^2_\mathcal{I}\,\mathcal{I}_{\rm end}^2 \approx \frac{1}{2}m^2_\mathcal{I}\,f^2 \approx \frac{1}{2}\lambda_{\mathcal{I}\phi} f^4$. Requiring $\rho_{\rm d}(t_{\rm end}) < \rho_\mathcal{I}(t_{\rm end})$, we have a lower bound on the Hubble parameter during inflation
\beqa
\label{eq:HI-lower}
H_{\mathcal{\cI}} \gtrsim H_{\mathcal{\cI}}^{\rm low} \equiv \frac{1}{\sqrt{6}}\, \lambda_{\mathcal{I}\phi}^{1/2}\,q_0^{1/3}\,f ~.
\eeqa
Similarly, requiring the dark monopole energy density to be less than the total dark sector energy density, or $\rho_{\tiny \Mcirc}(t_{\rm end}) = \Mmono \, n_{\tiny \Mcirc}(t_{\rm end}) < \rho_{\rm d}(t_{\rm end})$, $H_{\mathcal{I}}$ is also bounded from above~\footnote{Putting (\ref{eq:HI-lower}) and (\ref{eq:HI-high}) together gives $\lambda_{\mathcal{I} \phi} < q_0^{1/2} g^{2}$, which is less stringent than (\ref{eq:broken-before-inflation}).}
\beqa
\label{eq:HI-high}
H_{\mathcal{I}}\lesssim H_{\mathcal{I}}^{\rm high} \equiv 0.14\,q_0^{11/24}\, g^{1/2}\,\lambda_{\mathcal{I}\phi}^{1/4} \, f ~.
\eeqa

The third condition is that for the timeline in Fig.~\ref{fig:timeline} to hold, reheating cannot occur until after preheating has ended, \ie, $t_\text{end} < t_\text{RH}$.  This is the condition in (\ref{eq:aend-on-aRH}), or equivalently $\rho_\mathcal{I}(t_\text{end}) > \rho_\text{RH}$, which gives
\beqa
\label{eq:reheat-after-PR}
\frac{1}{2}\lambda_{\mathcal{I}\phi} f^4 > \frac{\pi^2}{30} g_{*,\text{RH}} T_\text{RH}^4 ~ .
\eeqa

The fourth general condition comes from the contribution to $\Delta N_{\rm eff}$ from the radiation degrees of freedom in the dark sector. The calculation can be seen from the schematic timeline in Fig.~\ref{fig:timeline}. At $t_{\rm end}$, the total energy in the dark sector is a (suppressed) fraction of the total inflaton energy. From $t_{\rm end}$ to $t_{\rm RH}$, the inflaton energy density scales like $a^{-3}$, while the dark sector radiation energy scales like $a^{-4}$. As a result, the dark sector radiation energy has an additional suppression factor of $a_{\rm end}/a_{\rm RH}$ compared to the visible sector energy $\rho_{\rm R}(t_{\rm RH}) \simeq \rho_{\mathcal{I}}(t_{\rm RH}) $. In terms of the model parameters, the contribution to the effective neutrino species or $\Delta N_{\rm eff}$ is
\beqa
\label{eq:DNeff}
\Delta N_{\rm eff} \approx 
\frac{8}{7}\left( \frac{11}{4} \right)^{4/3}\, \frac{\rho_d(t_{\rm end})}{2\,g^{-1}_{*\text{RH}}\,\rho_{\mathcal{I}}(t_{\rm end})}\,\frac{a_{\rm end}}{a_{\rm RH}} 
\approx
\left\{
\begin{array}{cc}
1.9\,\left( \frac{g_{*\mathrm{RH}}\,T_{\rm RH} }{\lambda_{\mathcal{I}\phi}^{1/4}\,f} \right)^{4/3} ~,   &  H_\mathcal{I} = H_\mathcal{I}^{\rm low}   \\
16\,\frac{\lambda_{\mathcal{I}\phi}^{1/2}}{g\,q_0^{1/4}}\,\left( \frac{g_{*\mathrm{RH}}\,T_{\rm RH} }{\lambda_{\mathcal{I}\phi}^{1/4}\,f} \right)^{4/3}  ~,  &  H_\mathcal{I} = H_\mathcal{I}^{\rm high}    
\end{array}
\right.
 ~. 
\eeqa
We note that the ratio in the parenthesis coincides with the condition $t_\text{end} < t_\text{RH}$ in (\ref{eq:aend-on-aRH}) and \eqref{eq:reheat-after-PR} up to a factor of $g_{*\text{RH}}$, so $\Delta N_\text{eff}$ ends up giving a slightly stronger constraint  than \eqref{eq:reheat-after-PR} when  $H_\mathcal{I} = H_\mathcal{I}^{\rm low}$ (see Sec.~\ref{sec:parameter-space-summary}), but not when $H_\mathcal{I} = H_\mathcal{I}^{\rm high}$ because $\lambda_{\cI\phi}^{1/2}/g$ must be suppressed according to (\ref{eq:abundance-preheat}).

The final general condition is to have the dark charged particle, $W'^\pm$, be non-relativistic at the end of parametric resonance, which could reduce the plasma effects and the monopole-antimonopole annihilation rate. We express the $W'^\pm$ density as $\rho_{W'}(t_{\rm end}) = \zeta \, \rho_{\rm d}(t_{\rm end})$, defining a dimensionless parameter $\zeta < 1$, which could be order of unity or suppressed, depending on the detailed dynamics of field configurations during and immediately after preheating. In our study, we will conservatively take $\zeta$ to be order of unity, which gives the highest likelihood that monopole-antimonopole annihilations could be important. Requiring $T'_\text{end} \simeq \rho_{W'}(t_{\rm end})^{1/4} < M_{W'} = g f$, one has 
\beqa
\lambda_{\mathcal{I}\phi}^{1/2} \lesssim
\bigg\{
\begin{array}{cc}
1.4 \, g^2 ~,   &  H_\mathcal{I} = H_\mathcal{I}^{\rm low}   \\
0.62 \, g^{5/3}\,q_0^{1/12}  ~,  &  H_\mathcal{I} = H_\mathcal{I}^{\rm high}    
\end{array}
  ~.
\eeqa

To study the plasma effects of $W'^\pm$ on the monopole annihilation rates, we will consider both cases with a thermalized and non-thermalized dark sector. We will show that the non-thermalized case can have dark monopoles with macroscopic scales.

%
\subsubsection{Thermalized dark sector}
\label{sec:thermalized}

We first consider the case with a kinetically thermalized dark sector. The temperature in the dark sector at $t_{\rm end}$ is then $T'_{\rm end} \simeq \rho_{d}(t_{\rm end})^{1/4}$, where we ignore the order-one factor for the radiation degrees of freedom in the dark sector. To keep a thermalized dark sector, the scattering rates should be higher than the Hubble scale at this time or $H(t_{\rm end}) \simeq [\rho_\mathcal{I}(t_{\rm end})]^{1/2}/M_{\rm pl}$. The scattering process $W' + \gamma' \rightarrow W' + \gamma'$ has an interaction rate of $\sigma_{\rm scat}v\,n_{\gamma'} \simeq g^3\,T'^2_{\rm end}/(4\pi\,f)$ at $t_{\rm end}$. Another relevant process to keep dark monopoles equilibrium is ${\scriptsize \Mcirc} + W' \rightarrow {\scriptsize  \Mcirc} + W'$ by exchanging a dark photon in $t$-channel. The corresponding interaction rate is $\sigma_{\rm scat}v\,n_{W'} \simeq 4\pi T'^2_{\rm end} / (g f)$ with $n_{W'}=\rho_{W'}/(g f)$, faster than the previous process. Requiring the slower interaction rate to be greater than $H(t_{\rm end})$, 
\beqa
\label{eq:kinetical-equi}
g^3 > 
\bigg\{
\begin{array}{cc}
1.7\,\pi^{5/2}\,\frac{f}{M_{\rm pl}} ~,   &  H_\mathcal{I} = H_\mathcal{I}^{\rm low}   \\
1.0\,\pi^{12/5}\,\lambda_{\mathcal{I}\phi}^{-3/10}\,q_0^{3/20}\, \left(\frac{f}{M_{\rm pl}}\right)^{6/5}  ~,  &  H_\mathcal{I} = H_\mathcal{I}^{\rm high}    
\end{array}
  ~.
\eeqa
Given the fact that $\sigma_{\rm scat} v\,n_{\gamma'} \propto a^{-2}$ and $H \propto a^{-3/2}$ during matter dominance, the interaction rate stays higher than the Hubble rate once the above condition is satisfied. 

For the plasma effects from $W'^\pm$, one needs to estimate the $W'^\pm$ abundance first. There are two situations, depending on whether $W'^\pm$'s reach chemical equilibrium or not. The annihilation rate for $W'^+ W'^- \rightarrow 2 \gamma'$ at $t_{\rm end}$ is estimated to be $\sigma_{\rm anni} v \simeq g^2/(4\pi\,f^2)$. Requiring $\sigma_{\rm anni} v \,n_{W'}(t_{\rm end}) > H(t_{\rm end})$, the condition for having chemical equilibrium is
\beqa
\label{eq:chemical-I}
\renewcommand{\arraystretch}{1.5}
\bigg\{
\begin{array}{cc}
g\,\lambda_{\mathcal{I}\phi}^{1/2} \gtrsim 8.9\,\pi^{3/2}\, \frac{f}{M_{\rm pl}} ~,   &\quad  H_\mathcal{I} = H_\mathcal{I}^{\rm low}  \\  
\lambda_{\mathcal{I}\phi} \gtrsim 3.5 \, \pi^{1/2}\,q_0^{1/4}\,\frac{f}{M_{\rm pl}}  ~,  &\quad  H_\mathcal{I} = H_\mathcal{I}^{\rm high}    
\end{array}
~.
\eeqa
Although the initial $W'$ abundance could be very high, it quickly reaches a thermal freeze-out value. Solving the Boltzmann equation, the freeze-out time is found to be $t_{\rm F} \simeq \frac{5}{2}\,t_{\rm end}$ with the freeze-out number density as
\beqa
n_{W'}(t_F)  \simeq \frac{\sqrt{\rho_{\mathcal{I}}(t_{\rm end})} }{ \sigma_{\rm anni} v \, M_{\rm pl}} \simeq \frac{2.8\,\pi \,\lambda_{\mathcal{I}\phi}^{1/2}   \, f^4}{g^2\,M_{\rm pl} }  ~.
\eeqa

Given the existence of a potentially large abundance of $W'^\pm$, we can estimate the free-streaming and capture lengths for formation of monopole-antimonopole bound states similarly to the Kibble-Zurek case. Using the monopole scattering cross section with the $W'$ in the plasma, $\sigma_{\rm scat} \simeq 4\pi/T'^2$, the free-streaming length is estimated to be
\beqa
\ell_{\rm free} \simeq \left( \frac{T'}{M_{\tiny \Mcirc}} \right)^{1/2}\, \frac{M_{\tiny \Mcirc}}{T'\,n_{W'}\, \sigma_{\rm scat}} \,, \qquad \qquad 
\ell_{\rm capt} \simeq \frac{4\pi}{g^2\, T'} ~.
\eeqa
At $t_F \simeq t_{\rm end}$ and using $H_\mathcal{I} = H_\mathcal{I}^{\rm low}$, the requirement of $\ell_{\rm free} > \ell_{\rm capt}$ becomes
\beqa
\label{eq:chemical-II}
\renewcommand{\arraystretch}{1.5}
\bigg\{
\begin{array}{cc}
g^{7/2}\,\lambda_{\mathcal{I}\phi}^{1/8} \gtrsim 6.4\,\pi^{15/4}\,\frac{f}{M_{\rm pl}} ~,   &\quad  H_\mathcal{I} = H_\mathcal{I}^{\rm low}   \\
g^{23/8}\,\lambda_{\mathcal{I}\phi}^{7/16}\,q_0^{-5/32} \gtrsim 3.5\,\pi^{25/8}\,\frac{f}{M_{\rm pl}}  ~,  &\quad  H_\mathcal{I} = H_\mathcal{I}^{\rm high}    
\end{array}
 ~.
\eeqa
The combinations of \eqref{eq:abundance-preheat}\eqref{eq:kinetical-equi}\eqref{eq:chemical-I}\eqref{eq:chemical-II} for the case with both kinetic and chemical equilibrium impose a stringent upper constraint on the dark monopole radius, which will be summarized in Section~\ref{sec:parameter-space-summary}.

If the $W'$ is not in chemical equilibrium with the dark photon plasma or \eqref{eq:chemical-I} is not satisfied, one should use the larger number density $n_{W'}(t_{\rm end})$ rather than $n_{W'}(t_F)$ to estimate the free-streaming length. The requirement of $\ell_{\rm free} > \ell_{\rm capt}$ then becomes
\beqa
\label{eq:chemical-III}
\renewcommand{\arraystretch}{1.5}
g \gtrsim 
\bigg\{
\begin{array}{cc}
1.1\,\pi^{11/10}\,\lambda_{\mathcal{I}\phi}^{3/20}  ~,   &\quad  H_\mathcal{I} = H_\mathcal{I}^{\rm low}   \\
1.2\,\pi^{25/23}\,\lambda_{\mathcal{I}\phi}^{9/46} \, q_0^{-3/92}   ~,  &\quad  H_\mathcal{I} = H_\mathcal{I}^{\rm high}    
\end{array}
~.
\eeqa
The combinations of \eqref{eq:abundance-preheat}\eqref{eq:kinetical-equi}\eqref{eq:chemical-III} and the reverse of \eqref{eq:chemical-I} are needed to identify the preferred parameter space to have dark monopoles account for all dark matter.

%
\subsubsection{Non-thermalized dark sector}
\label{sec:non-thermalized}

If \eqref{eq:kinetical-equi} is not satisfied, or
\beqa
\label{eq:eq:kinetical-equi-not}g^3 < 
\bigg\{
\begin{array}{cc}
1.7\,\pi^{5/2}\,\frac{f}{M_{\rm pl}} ~,   &  H_\mathcal{I} = H_\mathcal{I}^{\rm low}   \\
1.0\,\pi^{12/5}\,\lambda_{\mathcal{I}\phi}^{-3/10}\,q_0^{3/20}\, \left(\frac{f}{M_{\rm pl}}\right)^{6/5}  ~,  &  H_\mathcal{I} = H_\mathcal{I}^{\rm high}    
\end{array}
  ~,
  \eeqa
the $W'^\pm$ will not be generated by the thermal plasma.  We further expect that the population of particles produced during preheating will be dominantly massless modes.  Thus, we expect the $W'^\pm$ fraction of the dark sector energy density to be $\zeta \ll 1$.  In this case, monopole-antimonopole annihilations will be suppressed due to the absence of a charged plasma.  

To further justify that one could ignore the plasma effects for the preheating mechanism when the dark sector is out of thermal equilibrium, we start with $\zeta \sim 1$ and argue that $W'^\pm$ can be captured efficiently by the monopole field configuration.  As we describe here, the initial abundance $W'^\pm$ at $t_{\rm end}$ could be dramatically reduced if they are captured by the monopole field configuration (see discussion about the bound states of monopoles and charged particles in Refs.~\cite{Sonoda:1984nv,Goodband:1996xh}). The attractive potential between $W'^\pm$ and the monopole via the monopole and magnetic-dipole interaction prefers $W'^\pm$ to stay inside the monopole field configuration. Using the characteristic size of the dark field configuration to estimate the geometric capture cross section, one has
\beqa
\sigma_{\rm capt}(t_{\rm end}) \simeq \pi\,(\ell^{W'}_{\rm capt})^2 \simeq \pi\, p_*^{-2}\, \simeq 2.1\,\pi\, q_0^{1/6}\, \lambda_{\mathcal{I}\phi}^{-1}\,f^{-2}~.
\eeqa
We also note that this $\ell^{W'}_{\rm capt}$ coincides with equalizing the monopole-magnetic-dipole force to the centrifugal force by using the $W'$ magnetic dipole moment~\cite{Holstein:2006wi}, or  $g_\mathrm{M}/\ell^3 \times (2 g)/(2\,\mwd ) \simeq \mwd \, v^2_{W'}/\ell$ after substituting $v_{W'} \approx p_*/\mwd$. The $W'$ capture rate by monopoles is then
\beqa
\label{eq:wprime-capture-rate}
\Gamma^{W'}_{\rm capt}(t_{\rm end}) \simeq \sigma_{\rm capt}\, v_{W'}\, n_{\tiny \Mcirc}  \simeq 0.48\,\pi\,q_0^{-1/6}\,\frac{\lambda_{\mathcal{I}\phi}}{g}\,f ~.
\eeqa
Requiring $\Gamma^{W'}_{\rm capt}(t_{\rm end}) > H(t_{\rm end})$, one needs to satisfy
\beqa
\label{eq:wprime-capture-condition}
\lambda_{\mathcal{I}\phi}^{1/2} \gtrsim 2.4\,\pi^{-1/2}\,q_0^{1/6}\,g\,\frac{f}{M_{\rm pl} } ~.
\eeqa
In general, this capture rate decreases faster than the Hubble scale. In the earlier time before the monopole field configuration decreases to its final size $\Rmono \simeq (g \, f)^{-1}$, one has $\ell^{W'}_{\rm capt} \propto a$ and $\Gamma^{W'}_{\rm capt} \propto a^{-2}$. Given $H \propto a^{-3/2}$, the freeze-out time is $a(t_{F})/a_{\rm end}\simeq q_0^{-1/3}\,\lambda_{\mathcal{I}\phi} M_{\rm pl}^2/(g^2 f^2)$, and the $W'$ number density at the freeze-out time is
\beqa
\label{eq:Wp-abundance}
 n_{W'}(t_F) \approx n_{W'}(t_{\rm end})\, e^{- 2\,\Gamma^{W'}_{\rm capt}(t_{\rm end})/H(t_{\rm end})} \simeq \zeta\,\frac{\rho_{\rm d}(t_{\rm end})}{g\,f} \,e^{- 2\,q_0^{-1/6}\,\lambda_{\mathcal{I}\phi}^{1/2}\,M_{\rm pl}/(g \,f)} ~.
\eeqa
So, once the condition in \eqref{eq:wprime-capture-condition} is well satisfied, the $W'$ abundance is exponentially suppressed. Then, the plasma effects for monopoles are also suppressed. We can then use \eqref{eq:abundance-preheat} as an estimation of monopole abundance for this case with the constraints in \eqref{eq:eq:kinetical-equi-not}\eqref{eq:wprime-capture-condition}.

%
\subsection{Dark monopole radii and masses}
\label{sec:parameter-space-summary}

\begin{figure}[tb!]
	\label{fig:allowedProduction}
	\begin{center}
		\includegraphics[width=0.6\textwidth]{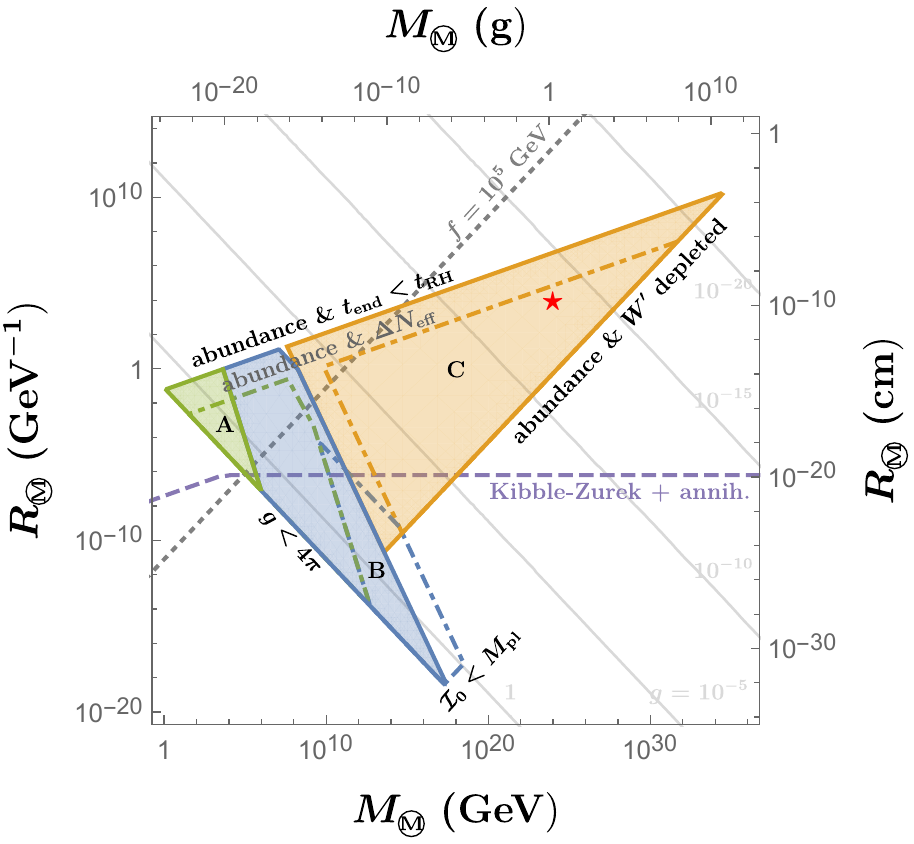}
		\caption{Regions (shaded) where dark monopoles can make up all of DM if produced via parametric resonance in $\Mmono \approx 4\pi f/g$ and $\Rmono \approx (gf)^{-1}$. Here, $T_\text{RH}=1~\text{MeV}$, $q_0=10^3$, and $\lambda=g^2$ have been assumed, marginalizing over all possible $\lambda_{\mathcal{I}\phi}$.  The shaded regions show $H_\cI=H_\cI^\text{high}$, while dot-dashed regions show $H_\cI^\text{low}$. In the orange ``C" region, the dark sector is out of both kinetic and chemical equilibrium. The blue ``B" region has only chemical equilibrium, while the green ``A" region has both chemical and kinetic equilibrium. The purple dashed line shows the result from the Kibble-Zurek production mechanism with a second-order thermal phase transition and subsequent dark monopole annihilations, with $\kappa\equiv T'/T=1/10$. Above this line, dark monopoles only account for a fraction of total DM.  The gray dotted line shows $f=10^5~\text{GeV}$; for smaller $f$ above this line, monopoles will not have EW-symmetric cores while maintaining $\lambda_{\phi h}$ small enough to decouple the dark sector from the SM (see text for details).  The light gray lines show contours of fixed $g$ varying from 1 to $10^{-20}$ from bottom to top in factors of $10^{-5}$.}
	\end{center}
\end{figure}

Figure \ref{fig:allowedProduction} shows the masses and radii where dark monopoles could make up all of DM, neglecting for the moment contributions to the DM density by other dark-sector particles.  Included are all three cases discussed in Sec.~\ref{sec:constraints}. We label these three different regions as ``A, B, C" with  A having both kinetic and chemical equilibrium for the dark sector, B having only kinetic equilibrium, and C out of both kinetic and chemical equilibrium. We have fixed $T_\text{RH} \simeq T_\text{BBN} \simeq 1~\text{MeV}$.  The plot is not very sensitive to our choice of $q_0 = 10^3$ (except the $\mathcal{I}_0 < \Mpl$ constraint).  We show $H_\mathcal{I} = H_\mathcal{I}^{\rm high}$ in the shaded regions surrounded by solid lines, whereas we show $H_\mathcal{I} = H_\mathcal{I}^{\rm low}$ in \eqref{eq:HI-lower} in the dot-dashed regions---there is very little difference between the two.

The boundaries for the plotted regions can be understood in the following way.  The boundary on the lower-left corner enforces gauge coupling perturbativity, $g<4 \pi$, using the relations for the monopole mass $\Mmono\approx 4 \pi f/g$ and radius $\Rmono \approx (gf)^{-1}$.  The $\mathcal{I}_0<\Mpl$ boundary comes from the choice of $q_0 = \mathcal{I}_0^2 / (4 f^2)$.  Notice that $f<\Mpl$ is always satisfied.  The kinetic equilibrium boundary (between regions ``B'' and ``C'') comes from (\ref{eq:kinetical-equi}).  The ``abundance \& $W'$ depleted'' condition comes from plugging the value for $\lambda_{\mathcal{I} \phi}$ in Eq.~(\ref{eq:wprime-capture-condition}) into the monopole abundance Eq.~(\ref{eq:abundance-preheat}). Below this, the $W'$ are not depleted and plasma-assisted monopole annihilations become important. Similarly, the ``abundance \& $t_\text{end} < t_\text{RH}$'' condition comes from plugging the inequality for $\lambda_{\mathcal{I} \phi}$ in Eq.~(\ref{eq:reheat-after-PR}) into Eq.~(\ref{eq:abundance-preheat}), and the ``abundance \& $\Delta N_\text{eff}$'' condition comes from using (\ref{eq:DNeff}) in (\ref{eq:abundance-preheat}).  
In (\ref{eq:DNeff}), we used $g_{*\text{RH}} \approx 10.75$ and the Planck 2018 result~\cite{Aghanim:2018eyx} $\Delta N_{\rm eff} < 0.28$ at 95\% confidence level. $\Delta N_\text{eff}$ dominates at large radius for $H_\cI^\text{low}$, while $t_\text{end}<t_\text{RH}$ dominates for $H_\cI^\text{high}$. Future CMB-S4 experiments~\cite{Abazajian:2016yjj} can improve this limit by one order of magnitude and probe the dark monopole parameter space from the large radius direction.
The boundary between regions ``A'' and ``B'' comes from substituting the condition for chemical equilibrium in Eq.~(\ref{eq:chemical-I}) into (\ref{eq:abundance-preheat}).
Finally, the upper boundary of region ``B'' comes from the requirement that monopoles do not annihilate in (\ref{eq:chemical-III}) substituted into (\ref{eq:abundance-preheat}) (this is most noticeable for $H_\cI = H_\cI^{\text{low}}$, but also explains the tiny notch for $H_\cI = H_\cI^{\text{high}}$).

These regions are compared against the predicted mass-radius relation if the monopoles are formed from a ``typical'' second-order phase transition (the Kibble-Zurek mechanism) followed by a period of plasma-assisted monopole annihilation and make up all of DM [the purple dashed line, given by the minimum of Eqs.~(\ref{eq:yieldKibbleZurek}) and (\ref{eq:yieldAnn})].  The region above this line would underproduce monopoles relative to the DM density.  For a perturbative gauge coupling $g < 4 \pi$, the plasma-induced monopole-antimonopole annihilations are always important, and the monopole radius is fixed as $\Rmono \simeq (g f)^{-1} \simeq 7 \times 10^{-7}~\text{GeV}^{-1}$ by Eq.~(\ref{eq:abundanceAnn}). Clearly, the parametric resonance production allows for a much wider range of radii to make up all of DM. One example point is shown with a red star in Fig.~\ref{fig:allowedProduction}.  This point corresponds to $\Mmono=10^{24}~\text{GeV} \approx 1.8\,\mbox{g}$, $\Rmono=10^4~\text{GeV}^{-1} \approx 2\times 10^{-10}\,\mbox{cm}$,  $f=2.8 \times 10^9~\text{GeV}$, $g=\sqrt{\lambda}=3.5 \times 10^{-14}$, and $\lambda_{\mathcal{I} \phi} = 2.6 \times 10^{-41}$, which has a very weakly interacting dark sector.  When using $H_\cI=H_\cI^{\text{low}}$, this point has $\Delta N_\text{eff} \approx 3.9 \times 10^{-2}$.

So far, we have only discussed the results based on the dark monopole abundance. The Higgs-portal coupling $\lambda_{\phi h}$ between the dark and SM sectors introduced in Section~\ref{sec:simpleHiggs} is constrained from above by requiring that the dark sector is not thermalized with the visible sector. For a reheating temperature below around 1 GeV, the scattering rate of the dark Higgs with SM light fermions is $\Gamma(T) \sim \sigma_{\rm scat}\,T^3 \sim \lambda_{\phi h}^2\,y_{f}^2\,v^2 \,T^3/(4\pi\,m_h^4)$. Requiring this rate to be slower than the Hubble rate, or $\Gamma \lesssim H \sim T^2/M_{\rm pl}$ at $T_{\rm RH}$, one needs to have $|\lambda_{\phi h}| \lesssim 10^{-5}$ using $T_{\rm RH}\sim 1$~GeV and the muon Yukawa coupling $y_\mu$.  For a lower reheating temperature below the muon mass, the electron Yukawa coupling should be used instead and the constraint on $\lambda_{\phi h}$ is weaker. To realize Case I in \eqref{eq:4-cases} with an EW-symmetric dark monopole core, $f > \sqrt{2}\,\mu_h/\sqrt{\lambda_{\phi h}} \gtrsim 10^5$~GeV, below the dotted gray line in Fig.~\ref{fig:allowedProduction}. Nevertheless, smaller $f$ is still allowed for other cases in \eqref{eq:4-cases}, though with less striking phenomenology because the monopole will not have an EW-symmetric core. Notice that only region ``C'' can produce monopoles with both an EW-symmetric core and large radius $\Rmono > 1/m_h$ as 100\% of DM.  
 
Our choice of $T_\text{RH} \simeq T_\text{BBN} \simeq 1~\text{MeV}$ makes the allowed regions in Fig.~\ref{fig:allowedProduction} as large as possible.  To see this explicitly, consider the example of region ``C'' when $H_\cI=H_\cI^\text{high}$.  The upper boundary comes from Eqs.~(\ref{eq:reheat-after-PR}) (lower line) and (\ref{eq:abundance-preheat}), leading to the relationship $T_\text{RH}^3 \Mmono^{-1/2} \Rmono^{3/2} = \text{constant}$ along the boundary.  The lower boundary comes from Eqs.~(\ref{eq:wprime-capture-condition}) and (\ref{eq:abundance-preheat}), leading to the relationship $T_\text{RH} \Mmono^{1/2} \Rmono^{-1/2} = \text{constant}$.  Thus, fixing $\Mmono$, an increase in $T_\text{RH}$ means the upper boundary decreases as $\Rmono \propto T_\text{RH}^{-2}$, while the lower boundary increases as $\Rmono \propto T_\text{RH}^2$.  Thus, the allowed parameter space disappears when $T_\text{RH}$ is too large.  More generally, to produce a monopole with radius larger than the radius obtained by Kibble-Zurek + annihilations in any of the regions ``A,B,C,'' $T_\text{RH} \lesssim 10~\GeV$ is required.

\section{Phenomenology}
\label{sec:pheno}

The main purpose of this paper to point out that the preheating production of dark monopoles could provide another type of macroscopic dark matter model. Furthermore, the interplay between the dark and EW gauge symmetry breaking via the simple Higgs-portal coupling can lead to an EW-symmetric core for the dark monopole. In this section, we briefly mention some phenomenological consequences of this dark monopole dark matter model. Some phenomenological consequences for an EW-symmetric dark matter ball can be found in Refs.~\cite{Ponton:2019hux,Bai:2019ogh}.

Other than the stable dark monopole, other stable particles in the dark sector may also maintain some relic abundance. For the simple $SU(2)/U(1)$ model here, additional particles are $h'$, $W'$, and $\gamma'$, all of which are required by the gauge group and are contained in the Lagrangian \eqref{eq:basic-dark-lag}. Similar to the SM Higgs boson, the dark Higgs is not a stable particle and can decay into two massless $\gamma'$. The $W'$ is stable if there is no lighter dark electric charged state in the model. For the case with a chemically self-thermalized dark sector and a large $W'$ mass, the ordinary thermal relic abundance of $W'$ could be much larger than the total DM energy density \cite{Khoze:2014woa}. A simple way to solve this problem is to introduce some massless dark charged fermions that $W'$ can decay to. Alternatively, the $W'$ can be captured efficiently by monopoles around the time of monopole formation.  This is the case, \eg, if (\ref{eq:wprime-capture-condition}) is well satisfied and the dark sector is out of kinetic equilibrium so that the $W'$ abundance is exponentially suppressed from \eqref{eq:Wp-abundance}. 

For Higgs-portal dark models, searching for the invisible Higgs decay at colliders could lead to the discovery of the dark sector. Using the experimental constraint on the invisible decay width $\Gamma_{\rm inv}/\Gamma_{\rm total} < 0.13$ at 95\% confidence level from ATLAS~\cite{ATLAS:2020cjb}, the Higgs-portal coupling is constrained to be $|\lambda_{\phi h}| < 3.0 \times 10^{-3}$ using $\Gamma(h \rightarrow \Phi \Phi) = 3 \lambda_{\phi h}^2 v^2 / ( 8\pi m_h)$. Note that the constraint on $|\lambda_{\phi h}|$ from Higgs invisible decays is less stringent than the one from requiring a thermally decoupled dark sector. 

The direct detection of dark monopoles depends on the interplay of the dark $SU(2)$ and SM EW symmetry breaking, as well as the dark monopole radius. When the radius is much smaller than the EW scale, or $\Rmono \ll 1/m_h$, the monopole can be treated as a point-like particle. Its effective coupling to the Higgs boson does depend on the monopole profile or the corresponding Higgs profile. For Case I in \eqref{eq:4-cases} with EW-symmetric monopoles, we can approximate the Higgs profile around the monopole as a step function and work in the Born approximation to estimate the monopole-nucleon scattering cross section for $\Rmono \ll R_{\rm th}$. Here, $R_{\rm th} \approx \frac{\pi}{2}(2 m_A V_0)^{-1/2}\approx 2.1/A~\mbox{GeV}$ is the threshold radius to have a bound state for the scattering with $V_0 \approx A\,y_{hNN}v$, $A$ is the atomic mass number, and $y_{hNN}\approx 1.1\times 10^{-3}$. The approximate potential well for scattering off a nucleus is $V(r) = A\,y_{hNN}\,\big(h(r)-v\big)\approx - V_0\,\Theta(r - \Rmono)$. The Born-approximate elastic spin-independent (SI) nucleon-monopole scattering cross section is~\cite{Ponton:2019hux}
\beqa
\label{eq:sigma-elastic-Born}
\sigma^{\rm elastic}_{N{\tiny \Mcirc}} &\approx& \frac{\sigma^{\rm elastic}_{A{\tiny \Mcirc}}}{A^2} \,=\, 4\pi\,\frac{1}{A^2}\,\left| \frac{m_A}{q} \int^\infty_0\,dr\,r\,\sin{(qr)}\,V(r) \right|^2     \\
&\approx& \frac{16\pi}{9}\, m_N^2\,A^2\,y_{hNN}^2\,v^2\,\Rmono^6 \approx (2.5 \times 10^{-42}\,\mbox{cm}^2)\,\left(\frac{A}{131}\right)^2\,\left(\frac{\Rmono}{10^{-3}\,\mbox{GeV}}\right)^6 ~.
\label{eq:sigma-elastic-Born-2}
\eeqa
Note that $\sigma^{\rm elastic}_{A{\tiny \Mcirc}} \propto A^4$. Comparing to the Xenon1T bound~\cite{Aprile:2018dbl}, the ordinary SI elastic scattering cross section upper bound is
\beqa
\sigma^{\rm SI}_{N{\tiny \Mcirc}}(\Mmono) \lesssim (1.2 \times 10^{-42}\,\mbox{cm}^2)\,\left( \frac{\Mmono}{10^6\,\mbox{GeV}} \right) ~.
\eeqa

For a larger $\Rmono$ above the threshold radius to have a bound state in the scattering process, the cross section becomes larger and can saturate the unitarity bound. As the radius keeps growing, many bound states mediate the scattering and the nucleus scattering cross section saturates the geometric one, or $\sigma^{\rm elastic}_{A\,{\tiny \Mcirc}} \approx 2 \pi\, \Rmono^2$ independent of $A$. 
When $\sigma^\text{elastic}_{A {\tiny \Mcirc}} n_\text{det} L_\text{det} \gtrsim 1$, with $n_\text{det}$ and $L_\text{det}$ the number density of target nuclei and length of the detector, the traditional single-hit search strategy becomes invalid.  For XENON1T, this is $\sigma^\text{elastic}_{A {\tiny \Mcirc}} \gtrsim [(1.42 \times 10^{22}~\text{cm}^{-3})(1~\text{m})]^{-1} \simeq 7 \times 10^{-25}~\text{cm}^2$ or $\Rmono \gtrsim 40~\GeV^{-1}$.
One should look for multi-hit scattering events in a detector that may or may not be underground (see Refs.~\cite{Bramante:2018qbc,Bramante:2018tos,Ponton:2019hux} for detailed discussion).

\begin{figure}[tb!]
	\label{fig:projected-sensivity}
	\begin{center}
		\includegraphics[width=0.6\textwidth]{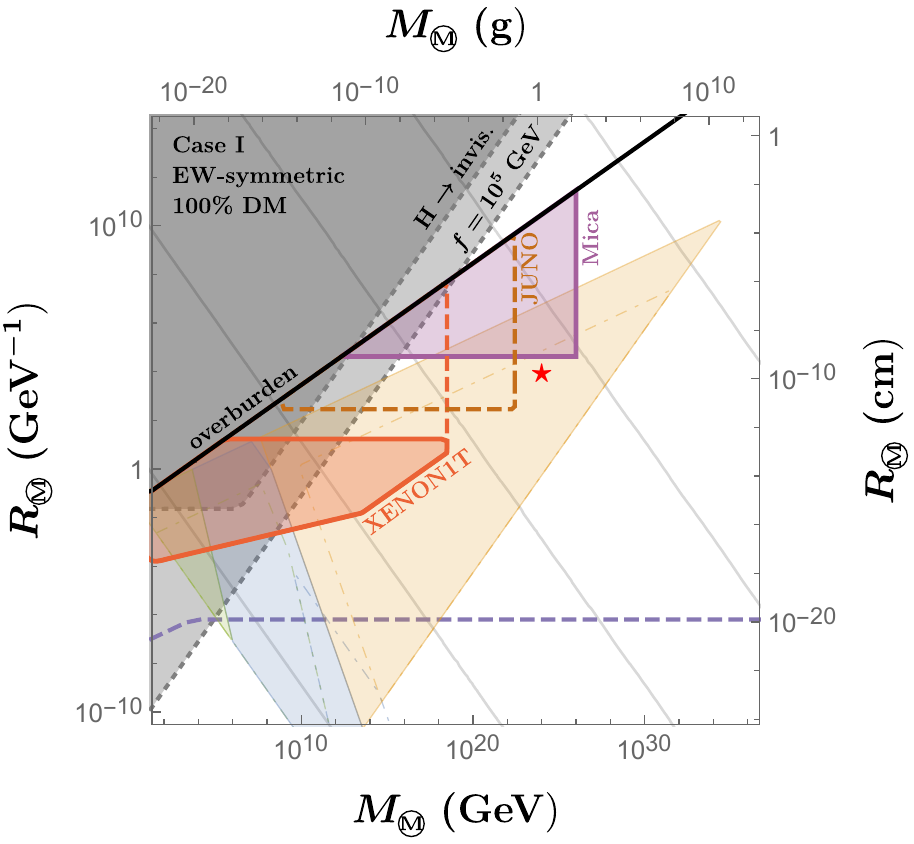}
		\caption{Constraints on monopoles with EW-symmetric cores, Case I above, assuming they make up 100\% of DM.  These are overlaid onto several features from Fig.~\ref{fig:allowedProduction}: the regions ``A,B,C,'' the $f=10^5~\GeV$ dotted gray line, the Kibble-Zurek$+$annihilations purple dashed line, and the light gray contours of $g$. The red shaded region is excluded by singlet-hit elastic scattering cross section constraints from  XENON1T~\cite{Aprile:2018dbl}.  The broader unshaded region within the red dashed line gives rise to multi-hit events at XENON1T, requiring a dedicated search.  The brown dashed line shows the JUNO \cite{An:2015jdp} sensitivity to multi-hit events \cite{Ponton:2019hux}.  The purple shaded constraint comes from the absence of signals in mica \cite{Price:1986ky,DeRujula:1984axn}.  The  black line shows where Earth and atmospheric overburden shielding prevents DM from making it to the detector.  The gray shaded regions above the dotted lines in the top left show where EW symmetry restoration in the core is impossible because of limits on $\lambda_{\phi h}$ from decoupling of the dark sector (lighter shading) and Higgs invisible decays \cite{ATLAS:2020cjb} with $\lambda_{\phi}=g^2$ (darker shading).}
	\end{center}
\end{figure}

The projected sensitivities in various experiments are shown in Fig.~\ref{fig:projected-sensivity}, overlaid on the features plotted in Fig.~\ref{fig:allowedProduction}. Here, dark monopoles are assumed to make up 100\% of DM at each point. 
In the shaded red region, direct detection experiments from XENON1T currently exclude monopole-nucleus elastic scattering \cite{Aprile:2018dbl}.  The break in the slope at about $\Mmono \simeq 10^{13}~\GeV$ and $\Rmono \simeq 10^{-2}~\GeV^{-1}$ corresponds to the change from the elastic scattering in (\ref{eq:sigma-elastic-Born-2}) to the resonance-mediated scattering cross section and eventually the geometric cross section (we simply take the geometric cross section here).  The dashed red region shows where multi-hit events are expected in these experiments.  This would require a dedicated search, as the signal would differ from elastic scattering.  The brown dashed region shows a similar multi-hit constraint for JUNO \cite{An:2015jdp}, assuming 10 years of taking data.  Present constraints on elastic scattering from mica \cite{Price:1986ky,DeRujula:1984axn} are shaded purple.  The upper-mass reach for all of these experiments is determined by the monopole number density, which must be large enough so that monopoles pass through the experiment during its data-taking.
The largest radius (or equivalently, the largest cross section) that can be probed is limited by DM deflections by Earth's atmosphere and crust above the detectors \cite{Kavanagh:2017cru,Bramante:2018qbc}.~\footnote{The underground detectors have all been taken to have the same depth $\sim 3~\text{km}$ for simplicity, giving a scattering cross section limit $\sigma \lesssim (2 \times 10^{-14}~\text{cm}^2) (\Mmono/10^{16}~\GeV)$ \cite{Ponton:2019hux}.}
As is evident in the plot, Earth shielding does not restrict the ability to probe EW-symmetric monopoles produced during preheating.
The Higgs-portal dark monopole model serves as an excellent motivation for those new avenues to search for dark matter.~\footnote{We have not shown constraints from direct detection surface runs \cite{Abusaidi:2000wg,Abrams:2002nb,Angloher:2017sxg,Davis:2017noy,Kavanagh:2017cru} and high-altitude detectors \cite{Rich:1987st,Starkman:1990nj,Wandelt:2000ad,Erickcek:2007jv,Mack:2007xj}, which are generally in the gray excluded regions above the underground overburden line. Indirect detection, IceCube \cite{Albuquerque:2010bt}, and Earth heating via captured DM annihilations \cite{Mack:2007xj, Bramante:2019fhi} would not provide constraints because dark monopoles annihilate primarily to the dark sector.}
In the gray shaded regions, $\lambda_{\phi h} f^2 < 2\,\mu_h^2$ is too small to restore the EW symmetry inside the monopole.
The light gray region above the dotted line (top-left) indicates where $f<10^5~\GeV$, corresponding to the bound $\lambda_{\phi h} \lesssim 10^{-5}$ coming from dark sector decoupling.  The dark gray region is bounded by Higgs invisible decays.  The smallest excluded radius depends on our choice $\lambda_\phi=g^2$;  for this choice, the bound cuts off at $m_h>2 \, m_{h'}= 2 \sqrt{2}\,g \, f = 2\sqrt{2} \Rmono^{-1}$ (no other constraints in this plot depend on $\lambda_\phi$).  

For Cases II and III in (\ref{eq:4-cases}) with the radius shorter than the weak scale, one can treat the dark monopole as a point-like particle. Similar to the SM particles that have their couplings to the Higgs boson proportional to their masses, one can use the variation of the monopole mass-squared in terms of EW VEV to derive the effective monopole-antimonopole-Higgs coupling 
\beqa
\mu_{h{\tiny \Mcirc}{\tiny \Mcirc}} = \frac{\partial \Mmono^2}{\partial v} = \left(\frac{4\pi\,Y}{g}\right)^2 \, \frac{\partial \Big( f^2 + \lambda_{\phi h} v^2/(2\,\lambda_\phi)\Big)  }{\partial v} = \left(\frac{4\pi\,Y}{g}\right)^2\,\frac{\lambda_{\phi h}}{\lambda_\phi}\,v ~,
\eeqa
with $Y=Y(\lambda_\phi/g^2)$ defined in (\ref{eq:Mmono}).~\footnote{One cannot simply take the monopole coupling to be the Higgs coupling to the $\Phi$ particle, which has been done in Ref.~\cite{Baek:2013dwa}. The monopole is a coherent state of $\Phi$ and gauge boson fields, so its coupling to the Higgs boson at a ``long" range is enhanced.} The SI nucleon-monopole scattering cross section is calculated as
\beqa
\label{eq:signa-elastic-effective}
\sigma^{\rm elastic}_{N{\tiny \Mcirc}} &\approx& \frac{\sigma^{\rm elastic}_{A{\tiny \Mcirc}}}{A^2} =\frac{m_N^2}{4\pi\,\Mmono^2}\,\frac{A^2 \, y^2_{hNN}\,\mu^2_{h{\tiny \Mcirc}{\tiny \Mcirc}}}{m_h^4} = 4\pi\,A^2\,\frac{\lambda_{\phi h}^2\,y^2_{hNN}\,Y^2}{g^2\,\lambda_\phi^2}\, \frac{m_N^2\,v^2}{f^2\,m_h^4} ~. 
\eeqa
Note that for $\lambda_\phi \approx g^2$ and $\Rmono \approx (g\,f)^{-1}$, the above formula is also proportional to $\Rmono^6$ and has the same dependence as in \eqref{eq:sigma-elastic-Born-2}.  Thus, compared to Case I, the elastic scattering cross section is suppressed by $(g^2/\lambda_\phi)^2 (\sqrt{\lambda_{\phi h}}\,f / m_h)^4 = (g^2/\lambda_\phi)^2 (\Delta v/ v)^4$, with $\Delta v$ the change of the EW vacuum in the center of the monopole compared to the exterior. Numerically, we have confirmed that the two ways to calculate the elastic scattering cross sections in Eqs.~\eqref{eq:sigma-elastic-Born} and \eqref{eq:signa-elastic-effective} agree with each other  when $\Rmono < 1/m_h$.

Other than elastic scattering, additional signatures come from the energetic photon emission from radiative capture of nuclei by the EW-symmetric dark monopole~\cite{Bai:2019ogh}. The emitted photons could be detected by detectors with higher threshold energies like Super-K, Hyper-K, or DUNE. Ref.~\cite{Bai:2019ogh} gives a conservative estimate for the radiative capture cross section in the large radius limit, which is several orders of magnitude less than the geometric cross section.  Using this very naive estimate, constraints from elastic scattering in mica supersede projected constraints from radiative captures in large volume experiments.  However, it is evident in their Fig.~5 that some modes have capture cross sections exceeding the geometric cross section.  Further dedicated study is necessary.

The bounds in Fig.~\ref{fig:projected-sensivity} are so strong that they could probe EW-symmetric monopoles as a small fraction of DM.  One may wonder whether they could measure monopoles with abundance fixed by annihilations after a thermal phase transition (\ie, ``Kibble-Zurek$+$annihilations''), as in \eqref{eq:abundanceAnn}.  For example, a radius $\Rmono \simeq 10^{-1}~\GeV^{-1}$ would correspond to $\Omega_{\tiny \Mcirc}/\Omega_{\rm dm} \simeq 10^{-10}$, which can actually be probed by XENON1T.  However, the maximum excluded mass is reduced by the DM density fraction.  Thus, only points inside the regime where the dark and visible sectors are coupled down to BBN can be excluded by XENON1T, and these points are already excluded by constraints on dark radiation degrees of freedom.

\section{Discussion and conclusions}
\label{sec:conclusion}

The interplays of the dark Higgs and SM Higgs profiles can induce an electroweak-symmetric core inside a monopole. So far, we have not mentioned the quantum chromodynamics (QCD) vacuum. Including QCD, the EW symmetry at the dark monopole core is at least broken by quark-antiquark condensation with a smaller induced VEV, $\langle H \rangle \sim \Lambda_{\rm QCD}$, much below the weak scale. This is a small correction to the scattering cross sections calculated before, but it may have nontrivial effects for the cosmological consequence of dark monopoles. For instance, the EW-symmetric dark monopole may induce baryon number violating processes~\cite{Arnold:1987mh}, which should happen before the QCD phase transition, {\eg}, if it is to explain baryogenesis.

Besides the renormalizable Higgs-portal operator, kinetic mixing between the dark and SM photons from a higher-dimensional operator could provide a coupling between these sectors.  Note that this is necessarily suppressed because the dark photon comes from a spontaneously broken non-Abelian gauge group [in our example, $SU(2)$].  Thus, the Higgs-portal coupling to the EW-modified monopole dominates for phenomenological purposes.

We also want to emphasize the features of using the preheating mechanism to produce dark monopoles or other topological defects in the early Universe.  If the dark sector is very weakly interacting, it could be non-thermalized, and the usual Kibble-Zurek topological defect production mechanism is inapplicable. A non-thermal production mechanism like preheating  can more efficiently shift the energy from the inflaton to the dark sector in order to produce topological defects.  It is a happy coincidence that existing and future experiments can already probe EW-symmetric dark monopoles produced in this way.  We also note that for a tiny dark gauge coupling $g$, the magnetic charge $4\pi/g$ could be very large. If dark monopoles are eaten by primordial black holes, the final primordial (nearly-)extremal black holes after Hawking radiation could have a large mass and provide another potentially interesting DM candidate~\cite{Bai:2019zcd,Maldacena:2020skw}.

In summary, dark monopoles in theories with small gauge couplings are an interesting macroscopic DM candidate. A Higgs-portal interaction to the dark gauge-symmetry-breaking field can lead to 't Hooft-Polyakov monopoles whose cores have a different Higgs vacuum expectation value.  Certain parameter choices even admit EW-symmetric cores. The conventional Kibble-Zurek production mechanism can produce large-radius monopoles with $\Rmono > 1/v$ at the expense of reducing the abundance due to monopole-antimonopole annihilations catalyzed by plasma effects. On the other hand, the parametric resonance production mechanism due to the coupling of the non-thermalized dark sector to the inflaton opens up a large region of $\Rmono -\Mmono$ parameter space that can still account for $100\%$ of dark matter. A large EW-symmetric core leads to an enhanced geometric scattering cross-section with nuclei. Current direct detection experiments exclude some regions of parameter space, while current and future experiments with the ability to search for multi-hit events could probe even larger radius and mass.

\subsubsection*{Acknowledgements}
We thank Joshua Berger and Akikazu Hashimoto for useful discussion.  
The work is supported by the U. S. Department of Energy under the contract DE-SC0017647. Part of this work was performed at the Aspen Center for Physics and KITP, which are supported by National Science Foundation grants, NSF PHY-1066293 and NSF PHY-1748958, respectively.  

\setlength{\bibsep}{6pt}
\bibliographystyle{JHEP}
\bibliography{mono_higgs_refs}

\end{document}